\documentclass[twocolumn]{aastex62}
\usepackage{xcolor}
\usepackage{url}
\usepackage{amsmath}
\newcommand{\package}[1]{\textsl{#1}}

\newcommand{\kms}{\mbox{km s$^{-1}~$}} 
\newcommand{\kmse}{\mbox{km s$^{-1}$}} 

\newcommand{\kmsmyre}{\mbox{km s$^{-1}$ Myr$^{-1}$}}

\newcommand{\msune}{\ensuremath{\textrm{M}_{\odot}}} 
\newcommand{\msun}{\msune~} 
\newcommand{\vhelio}{\ensuremath{V_{\rm hel}}~}
\newcommand{\vhelioe}{\ensuremath{V_{\rm hel}}}
\newcommand{\vlsr}{\ensuremath{V_{\rm LSR}}~}
\newcommand{\vlsre}{\ensuremath{V_{\rm LSR}}}

\newcommand{\lmse}{\ensuremath{L_{\rm MS}}}
\newcommand{\lms}{\lmse~}
\newcommand{\bmse}{\ensuremath{B_{\rm MS}}}
\newcommand{\bms}{\bmse~}
\newcommand{\dgr}{$^{\circ}~$}

\newcommand{\nhi}{$N_{\rm H \small{I}}$ }

\newcommand{\hi}{H{\footnotesize I} }

\newcommand{\teff}{$\rm T_{eff}$~}
\newcommand{\teffe}{$\rm T_{eff}$}
\newcommand{\logg}{$\log{g}$~}
\newcommand{\logge}{$\log{g}$}

\newcommand{\pwi}{\textit{PW~1}~}
\newcommand{\pwie}{\textit{PW~1}}
\newcommand{\laii}{LA II~}
\newcommand{\laiie}{LA II}
\defcitealias{PriceWhelan2018}{Paper I}

\newcommand{\kpc}{\ensuremath{\textrm{kpc}}}
\newcommand{\gaia}{\textit{Gaia}}
\newcommand{\cannon}{\textit{Cannon}}
\newcommand{\thecannon}{\textit{The Cannon}}
\newcommand{\bs}[1]{\boldsymbol{#1}}
\renewcommand{\vec}[1]{\bs{#1}}
\newcommand{\mat}[1]{\mathbf{#1}}
\newcommand{\given}{\,|\,}
\newcommand{\masyr}{\ensuremath{\textrm{mas}\,\textrm{yr}^{-1}}}
\newcommand{\pwlong}{\textit{Price-Whelan~1}}

\newcommand\numberthis{\addtocounter{equation}{1}\tag{\theequation}}

\newcommand{\measvdisp}{11.0} 
\newcommand{\measvlsrk}{273.4} 
\newcommand{\measvr}{276.7} 

\shorttitle{\pwi Spectroscopy}
\shortauthors{Nidever et al.}

\begin{document}

\title{Spectroscopy of the Young Stellar Association \textit{Price-Whelan 1}: \\
Origin in the Magellanic Leading Arm and Constraints on the Milky Way Hot Halo}
\correspondingauthor{David Nidever}
\email{dnidever@montana.edu}

\author[0000-0002-1793-3689]{David Nidever}
\affiliation{Department of Physics, Montana State University, P.O. Box 173840, Bozeman, MT 59717-3840}
\affiliation{National Optical Astronomy Observatory, 950 North Cherry Ave, Tucson, AZ 85719}

\author[0000-0003-0872-7098]{Adrian M. Price-Whelan}
\affiliation{Center for Computational Astrophysics, Flatiron Institute, 162 Fifth Avenue, New York, NY 10010, USA}
\affiliation{Department of Astrophysical Sciences, Princeton University, 4 Ivy Lane, Princeton, NJ 08544, USA}

\author[0000-0003-1680-1884]{Yumi Choi}
\affiliation{Department of Physics, Montana State University, P.O. Box 173840, Bozeman, MT 59717-3840}
\affiliation{Space Telescope Science Institute, 3700 San Martin Drive, Baltimore, MD 21218, USA}

\author[0000-0002-1691-8217]{Rachael L.~Beaton}
\altaffiliation{Hubble Fellow}
\altaffiliation{Carnegie-Princeton Fellow}
\affiliation{Department of Astrophysical Sciences, Princeton University, 4 Ivy Lane, Princeton, NJ 08544, USA}
\affiliation{The Observatories of the Carnegie Institution for Science, 813 Santa Barbara St., Pasadena, CA~91101}

\author{Terese T.~Hansen}
\affiliation{Mitchell Institute for Fundamental Physics and Astronomy and Department of Physics and Astronomy, Texas A\&M University, College Station, TX~77843-4242, USA}

\author{Douglas Boubert}
\affiliation{Institute of Astronomy, Madingley Rd, Cambridge CB3 0HA, UK}

\author{David Aguado}
\affiliation{Institute of Astronomy, Madingley Rd, Cambridge CB3 0HA, UK}

\author{Rana Ezzeddine}
\affiliation{Department of Physics and Kavli Institute for Astrophysics and Space Research, Massachusetts Institute of Technology, Cambridge, MA 02139, USA}

\author[0000-0001-7790-5308]{Semyeong Oh}
\affiliation{Institute of Astronomy, Madingley Rd, Cambridge CB3 0HA, UK}

\author{N. Wyn Evans}
\affiliation{Institute of Astronomy, Madingley Rd, Cambridge CB3 0HA, UK}

\begin{abstract} 
We report spectroscopic measurements of stars in the recently discovered young stellar association \pwlong\ (\pwie), which was found in the vicinity of the Leading Arm (LA) of the Magellanic Stream. We obtained Magellan+MIKE high-resolution spectra of the 28 brightest stars in \pwi and used The Cannon to determine their stellar parameters. We find that the mean metallicity of \pwi is [Fe/H]=$-$1.23 with a small scatter of 0.06 dex and the mean radial velocity is $\vhelio=\measvr$ \kms with a dispersion of \measvdisp~\kmse. Our results are consistent in \teffe, \logge, and [Fe/H] with the young
and metal-poor characteristics (116 Myr and [Fe/H]=$-$1.1) determined for \pwi from our discovery paper. We find a strong correlation between the spatial pattern of the \pwi stars and the \laii gas with an offset of $-$10.15\dgr in \lms and $+$1.55\dgr in \bmse. The similarity in metallicity, velocity, and spatial patterns indicates that \pwi likely originated in \laiie. We find that the spatial and kinematic separation between \laii and \pwi can be explained by ram pressure from Milky Way gas.  Using orbit integrations that account for the LMC and MW halo and outer disk gas, we constrain the halo gas density at the orbital pericenter of \pwi to be $n_{\rm halo}(17~\kpc) = 2.7_{-2.0}^{+3.4} \times 10^{-3}~{\rm atoms}~{\rm cm}^{-3}$ and the disk gas density at the midplane at $20~\kpc$ to be $n_{\rm disk}(20~\kpc,0) = 6.0_{-2.0}^{+1.5} \times 10^{-2}~{\rm atoms}~{\rm cm}^{-3}$.
We, therefore, conclude that \pwi formed from the \laii of the Magellanic Stream, making it a powerful constraint on the Milky Way--Magellanic interaction.
\end{abstract}

\keywords{Galaxy: open clusters and associations – Galaxy: halo – stars: formation – surveys - Magellanic Clouds}


\section{Introduction} \label{sec:intro}

The gaseous Magellanic Stream (MS) and its Leading Arm (LA) component, which, respectively, trail and lead the Magellanic Clouds (MCs) in their orbit about the Milky Way (MW), are one of the most prominent \hi features in the sky \citep{WW1972,Mathewson1974,Putman1998,Putman2003,Bruens2005,Nidever2008,Stanimirovic2008}. Together, the MS and LA stretch over 200\dgr from end to end \citep{Nidever2010}, and are a prototypical example of a gaseous stream stripped from a satellite galaxy in the process of being accreted onto the MW.  

The LA is composed of four main complexes \citep[LA I--IV;][]{Bruens2005,For2013}, is shorter than the trailing stream, and has a more irregular shape, the latter two characteristics being likely due to the effects of ram pressure. In fact, many of the LA cloudlets show head-tail shape \citep[e.g.,][]{McClure-Griffiths2008}. Two of the components, LA II and III, lie above the Galactic plane (i.e., at positive Galactic latitudes), suggesting that these complexes have already passed through the Galactic midplane. The formation of the LA has been a topic of great debate, but the proposed formation mechanisms have generally broken down into three primary physical processes: tidal stripping  \citep[e.g.,][]{Gardiner1996,Connors2006,Besla2012,Diaz2012}, ram pressure stripping  \citep[e.g.,][]{Mastropietro2005}, and stellar feedback \citep[e.g.,][]{Olano2004,Nidever2008}.  
Each formation mechanism has its own strengths and weaknesses in terms of explaining the LA morphology and kinematics. In the end, the formation of the LA is likely a combination of all three of these mechanisms, albeit their relative importance and chronological sequence remains an active topic of research \citep[see the recent review by][]{Donghia2016}.  

With high-precision proper motion measurements of the MCs \citep[e.g.,][]{Kallivayalil2006,Besla2007,Kallivayalil2013,Helmi2018}, the first-infall scenario, e.g., that the MCs are on their first passage around the MW, has become a widely accepted model to explain both the dynamical evolution of the MCs and their interaction with the MW. 
In this scenario, the MCs were dynamically-bound long before they fell into the MW potential $\sim$1~Gyr ago \citep{Besla2012}.  
Many threads of observational evidence suggest that the MCs had the first strong gravitational interaction $\sim$2--3~Gyr ago  \citep[e.g.,][]{Harris2004,Harris&Zaritsky2009,Weisz2013} and then experienced a direct collision a few hundred Myr ago \citep[e.g.,][]{Olsen2011,Besla2012,Noel2015,Choi2018a,Choi2018b,Zivick2018}. 
The majority of gas in the LA and MS was likely stripped during the former interaction, while the gas in the Magellanic Bridge was likely stripped off during the latter. 
However, the exact timing, mass, and origin (e.g., LMC or SMC) for the gas required to form the LA, MS, and the Magellanic Bridge remain unknown \citep{Pardy2018}.

Due to the differences in the star formation and baryon cycle histories of the LMC and SMC, the chemical abundances of the gas provide critical clues to understanding the origin and evolution of the Magellanic system. 
Abundances for high velocity clouds associated with the MS have been measured using absorption along the line-of-sight to a bright background source (a quasar or hot star), but appropriate sight lines are limited in number and location to produce a global characterization. 
A number of {\em HST} studies have found that the MS has a mean metallicity of [Fe/H] $\approx$ $-$1 dex across its length \citep{Fox2010,Fox2013b}, although there is evidence for a more metal-rich component \citep{Richter2013,Fox2013b}.  The Magellanic Bridge, the gas between the MCs, also has a metallicity of $\approx$ $-$1 \citep{Lehner2008} as does the LA \citep{Lu1998,Fox2018,Richter2018}.  The consistency in the metallicities for these distinct gaseous features that lead, trail, and connect the MCs suggests that they share a common originating system: the MCs, themselves.  
The current peak stellar metallicities of the LMC and SMC are [Fe/H]=$-$0.6 dex and [Fe/H]=$-$1.0 dex, and even higher in the innermost regions, which have ongoing star formation, reaching [Fe/H]=$-$0.2 and [Fe/H]=$-$0.7, respectively \citep{Nidever2019b}.
The current mean metallicity of the gas in the LMC and SMC is also high --- [Fe/H]=$-$0.2 and [Fe/H]=$-$0.6, respectively \citep{Russell1992} --- consistent with the most metal-rich stars.
However, the gas components that formed the LA and MS were stripped some time ago and, therefore, their chemistry should be compared to the MCs in the past.
According to \citet{Pagel1998,Harris2004} the MC metallicities were $\sim$0.3 dex lower $\sim$2 Gyr ago making the SMC's metallicity consistent with those measured in the gaseous MS, LA, and Bridge today.

While the MCs are known to have extensive stellar peripheries \citep[e.g.,][]{Munoz2006,Majewski2009,Nidever2011,Nidever2019a}, no stars have been detected in the MS or LA despite many attempts
\citep[e.g.,][]{Philip1976a,Philip1976b,Recillas-Cruz1982,Bruek1983,Kunkel1997,Guhathakurta1998}. 
Recent star formation in the LA is expected because molecular hydrogen is ubiquitous in the LA \citep[e.g.,][]{Richter2018} and shock compression caused by gas in both the Galactic disk and halo is anticipated. 
Indeed, \citet{CasettiDinescu2014} discovered a number of young stars in the region of the LA that had radial velocities consistent with being formed in the LA gas, initially confirmed by follow-up, high-resolution spectroscopy with Magellan+MIKE that enabled chemical abundance measurements of these stars. 
However, more recent high-resolution spectroscopy and orbital analyses incorporating Gaia DR2 \citep{GaiaDR2} proper motions from the same team determined that the stars were incompatible with a Magellanic origin \citep{Zhang2019}. 

\begin{figure*} 
\begin{center}
\includegraphics[width=0.91\hsize,angle=0]{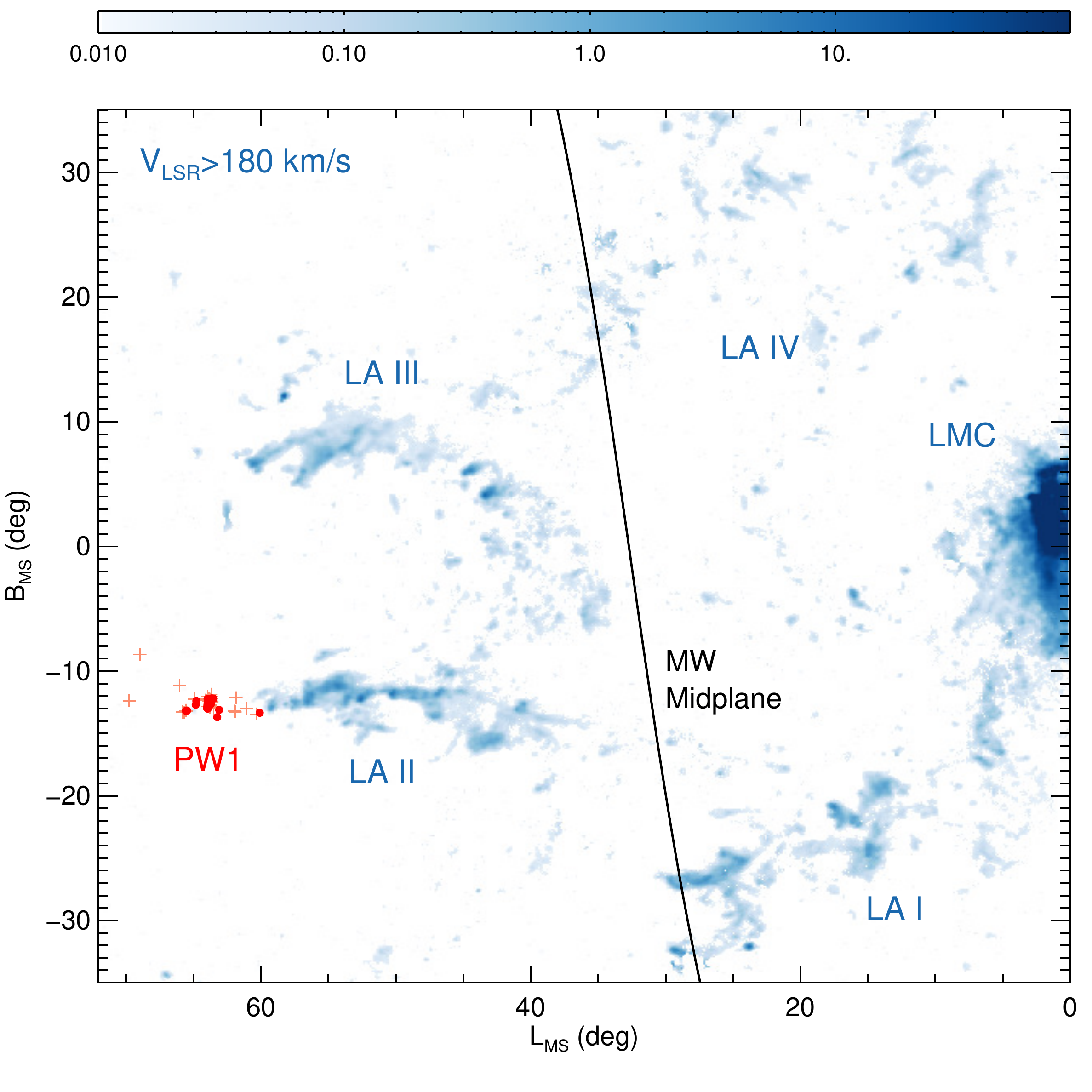}
\end{center}
\caption{Map of the Leading Arm region.  The GASS \hi column density (in units of 10$^{19}$ atoms cm$^{-2}$) is shown in blue with the LA I-IV and LMC labeled.  The 28 PW1 stars for which we obtained MIKE spectra shown as filled red circles while the light-red crosses are 43 additional PW1 candidates based on photometry and Gaia DR2 proper motions.  The MW midplane is shown by a solid black line.}
\label{fig_map}
\end{figure*}

While the initial claims of LA-associated young stars were invalidated with Gaia DR2 proper motions, \citet[][hereafter Paper I]{PriceWhelan2018} discovered a young stellar association (\textit{``Price-Whelan 1''}; hereafter \pwie) using \gaia\ proper motions in the vicinity of the LA. In \citetalias{PriceWhelan2018}, \pwi was found to be young, metal-poor, and likely disrupting with a sky position similar to that of LA II. By comparing to stellar evolution models, \citetalias{PriceWhelan2018} measured an age of $\sim$116~Myr, distance of $\sim$29~kpc, metallicity of [Fe/H] = $-$1.14, and total present-day stellar mass of $\sim$1200~\msune. Although the radial velocity (RV) of the cluster was unknown at the time, the range of possible orbits strongly suggested an association with the MCs and a passage through the outer MW disk $\sim$116 Myr ago.
However, there are many \hi structures in the vicinity of \pwie\ (Figure~\ref{fig_map}), each with a unique RV signature. 
Associating \pwie with a particular gaseous substructure therefore requires spectroscopic RV measurements of the cluster stars.


%
%


In this paper, we analyze the stellar parameters and mean kinematics of the stars in \pwi to better constrain the properties and origin of this young stellar association.
We also use the kinematics of \pwi in an orbital analysis to constrain the density of the MW hot halo and outer gas disk.
The layout of the paper is as follows: Section \ref{sec:data} describes the observations and data reduction.  In Section \ref{sec:analysis}, we present the procedures for deriving radial velocities and stellar parameters.  Our main results are detailed in Section \ref{sec:results} and discussed in Section \ref{sec:discussion}.  Finally, our main conclusions are summarized in Section \ref{sec:summary}.

\section{Observations and Reductions} \label{sec:data}

\begin{figure}[t]
\begin{center}
\includegraphics[width=1.0\hsize,angle=0]{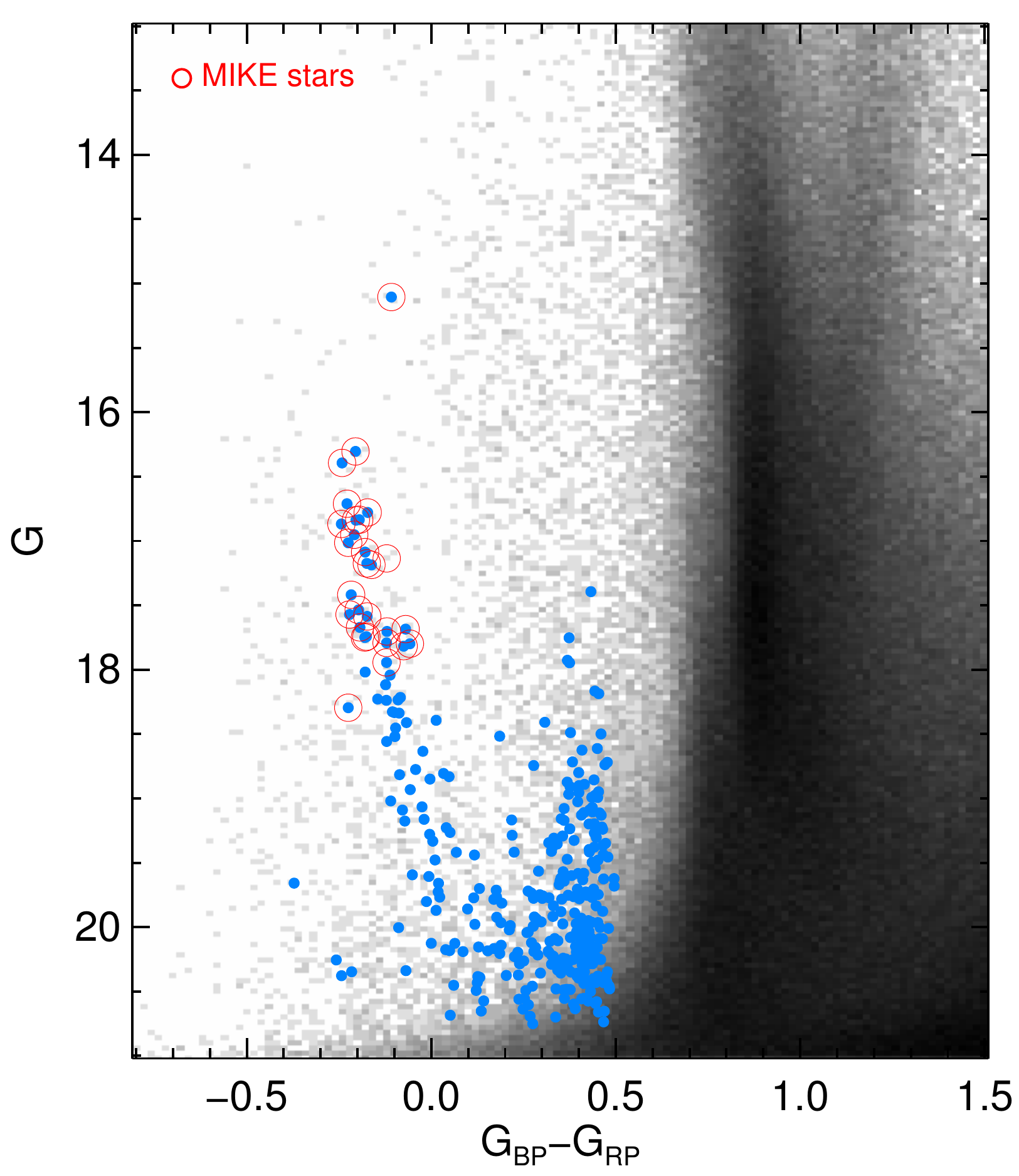}
\end{center}
\vspace{-0.5cm}
\caption{
Color-magnitude diagram using \gaia\ photometry in the vicinity of \pwie (black points); the specific spatial constraints are 173.5\dgr $<$ $\alpha$ $<$ 185.5\degr, $-$35.0\dgr $<$ $\delta$ $<$ $-$23.0\degr. 
Filled (blue) circles are probable \pwi members (with $G_{\rm BP} - G_{\rm RP} < 0.5$) based on the proper motion model described in \citetalias{PriceWhelan2018}.
MIKE spectra were obtained for the 28 stars indicated by open red circles, including PW1-03 which was not used in the analysis due to low S/N.}
\label{fig_cmd}
\end{figure}

We obtained spectra for the brightest 28 \pwi stars and 6 standard stars using the Magellan Inamori Kyocera Echelle (MIKE) spectrograph \citep{Bernstein2003} on Magellan-Clay at Las Campanas Observatory on 25, 26, and 30 April and May 1 2019. 
Some of the stars were also observed on 26, 27, 28 December 2018 using the Goodman Spectrograph \citep{Clemens2004} at the SOAR Telescope with consistent results.
Table 1 presents the names, coordinates, magnitudes, signal-to-noise ratio (S/N) and proper motions for each target.  \autoref{fig_cmd} shows their location in a \emph{Gaia} color-magnitude diagram (red open circles) relative to all \pwi candidates (blue) and field stars (black).
Observations were taken with the 0.7$''$ slit and 2x2 pixel binning, such that the native resolution for each spectrum is $R=31,000$. 
Exposure times ranged from 70 to 1060 seconds under generally good conditions with an average seeing of 0.7$''$.  
A the native MIKE resolution, S/N per resolution element (3.15 pixels) for the \pwi stars is 10--24 with a median of $\sim$18, although the S/N per pixel is increased by 3.16$\times$ by rebinning the 1-D extracted spectra (see below).
Six of the stars have low S/N due to moderately cloudy conditions (PW1-00 -- PW1-05) and are generally excluded from detailed analysis.
An additional high-S/N spectrum was obtained for PW1-00 (the brightest \pwi member) that is suitable for precise chemical abundance work, but this analysis is reserved for future work. 

\begin{figure*}[t]
\begin{center}
\includegraphics[width=1.0\hsize,angle=0]{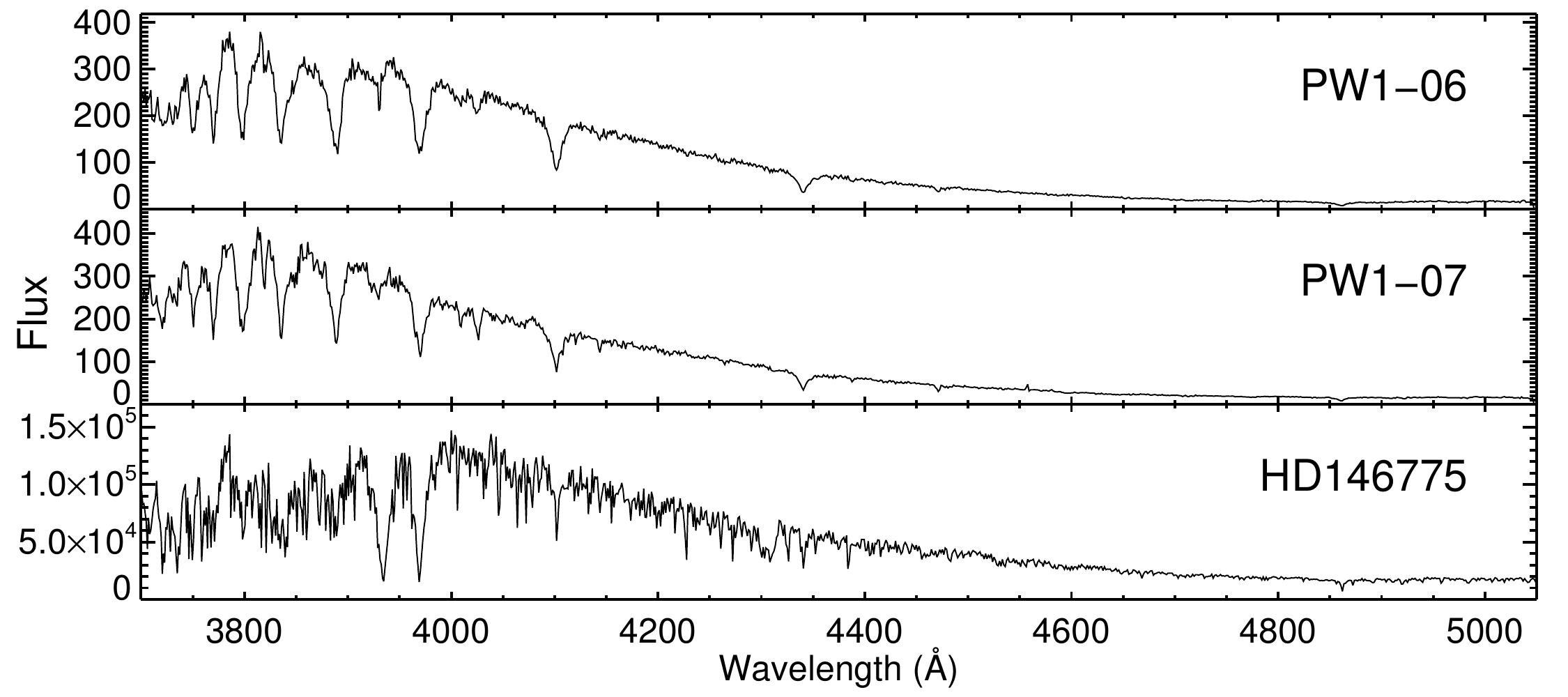}
\end{center}
\vspace{-0.5cm}
\caption{Three example MIKE blue-arm spectra.  
The top two panels show example \pwi stars, PW1-06 (s/N=29) and PW1-07 (S/N=30) , while the bottom panel is the reference star HD146775 (S/N=826).}
\label{fig_spectra}
\end{figure*}

The observations were processed using the MIKE pipeline\footnote{Available: \url{https://code.obs.carnegiescience.edu/mike}}, which is part of the CarPy spectroscopic reduction package that uses algorithms described in \citet{Kelson2000} and \citet{Kelson2003}. 
The MIKE pipeline performs image processing (bias removal, flat fielding) as well as extracting the spectra, determining a wavelength solution, subtracting the sky, and co-adding multiple exposures of single objects. The multi order reduced spectra were then merged into a single spectrum and normalized using the IRAF\footnote{IRAF is
distributed by the National Optical Astronomy Observatories, which are operated by the Association of Universities for Research in Astronomy, Inc., under cooperative agreement with the National Science Foundation.} tasks {\tt continuum} and {\tt scombine}.  While MIKE produces spectra from the blue and red arms, only blue spectra with a wavelength range of 3550--5060\AA\ were used in our analysis.
Figure~\ref{fig_spectra} shows example spectra for two \pwi stars and the standard star HD 146775.

Finally, the full MIKE resolution was significantly higher than needed for our science goals.  Therefore, we rebinned the spectra with a bin size of 10 native MIKE pixels which increased the S/N per pixel by $\sim$3.16$\times$.  This had little impact on the spectral features as they are already quite broad in these hot, B-type stars.


\section{Data Analysis} \label{sec:analysis}

\subsection{Radial Velocities} \label{subsec:rv}
A RV was determined for each star via a two phase process. 
First, we determine an initial RV for each source by cross-correlating the unsmoothed MIKE spectra (excluding the region below 3665\AA) with a hot synthetic stellar spectrum having stellar parameters (\teffe, \logge, [Fe/H]) = (15,000K, 4.0, $-$1.0) and set to the the resolution and logarithmic wavelength scale of the MIKE spectra. 
Second, the RVs were then redetermined using its best-fit model as the cross-correlation template after the star-by-star stellar parameters were determined following the procedure described in \S \ref{subsec:cannon}. The second iteration provided similar, but more precise and accurate RV solutions. 

We determine the RV uncertainties using a Monte Carlo scheme. Mock observations were generated for each star by adding Gaussian noise similar to that observed in the science spectrum to its best-fit model spectrum. One hundred mock observations were generated for each star, the RV determined for each with the method described above, and the uncertainty calculated as the robust standard deviation of these values. The typical RV uncertainty from this procedure is $\sim$3 \kms but the uncertainty increases roughly with the inverse of S/N. 

We compared our RVs for the six bright standard stars to literature values and find a median offset of $-$2.0 \kmse, which is consistent within the uncertainties of our measurements.



\begin{figure*}
\begin{center}
\includegraphics[width=0.90\hsize,angle=0]{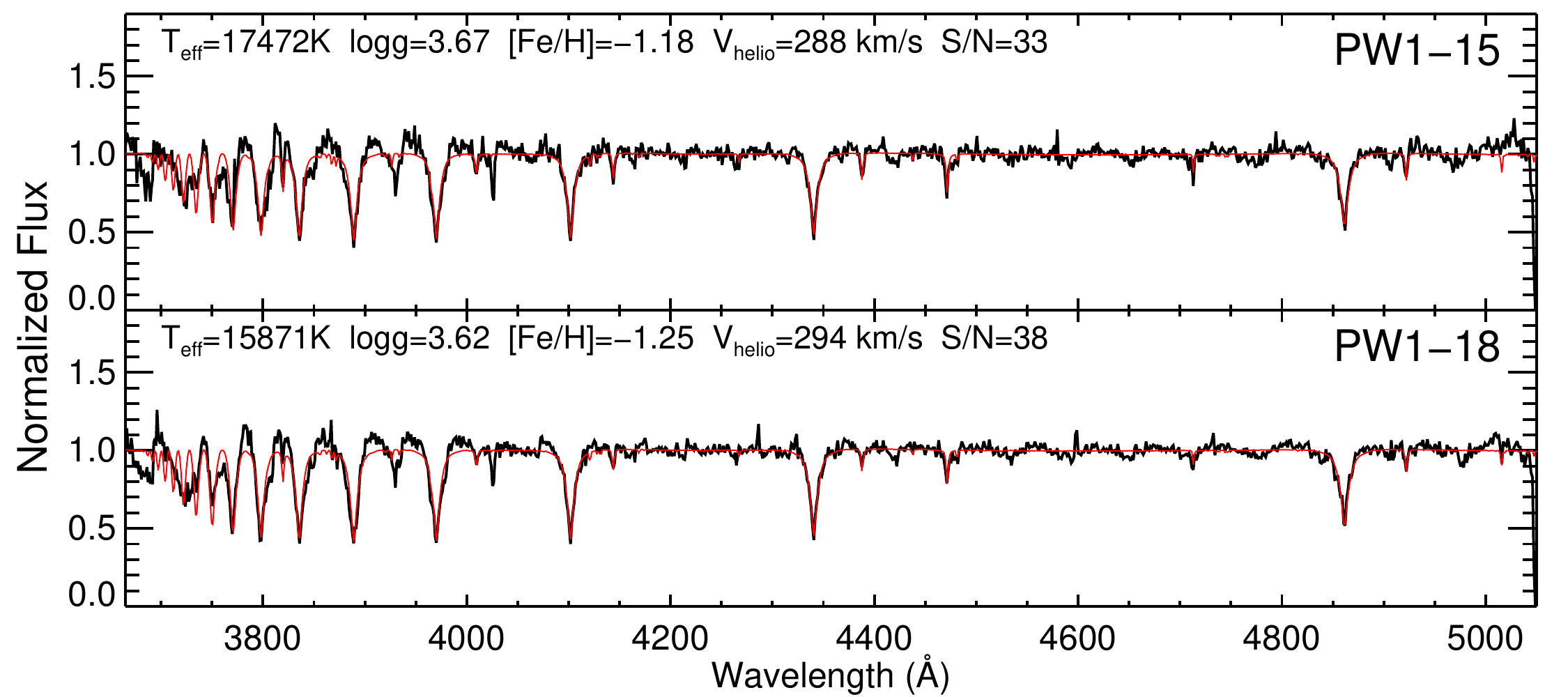}
\end{center}
\caption{Example best-fit spectra from \thecannon. MIKE spectra for PW1-15 and PW1-18 (black) are shown with the best-fit Cannon models overlaid (red). These show excellent agreement. The best-fit stellar and RV parameters are noted in the top left corner for each star. }
\label{fig_spectra_model}
\end{figure*}

\begin{figure}
\begin{center}
\includegraphics[width=1.0\hsize]{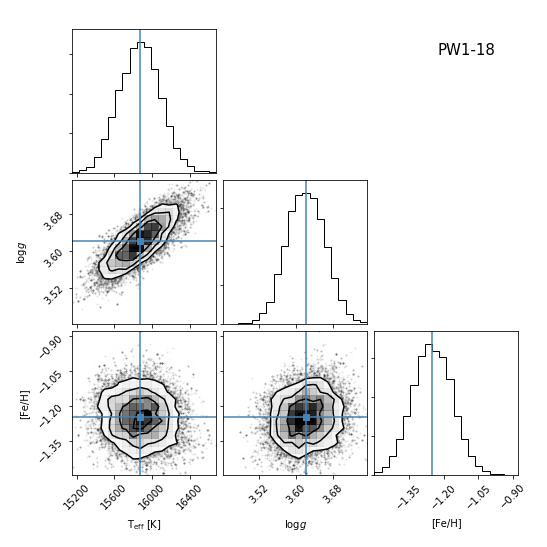}
\end{center}
\caption{An example corner plot showing the {\em emcee} MCMC posterior distributions for PW1-18.  The median values are shown by the blue lines.}
\label{fig_cannoncorner}
\end{figure}

\begin{figure} 
\begin{center}
\includegraphics[width=0.91\hsize,angle=0]{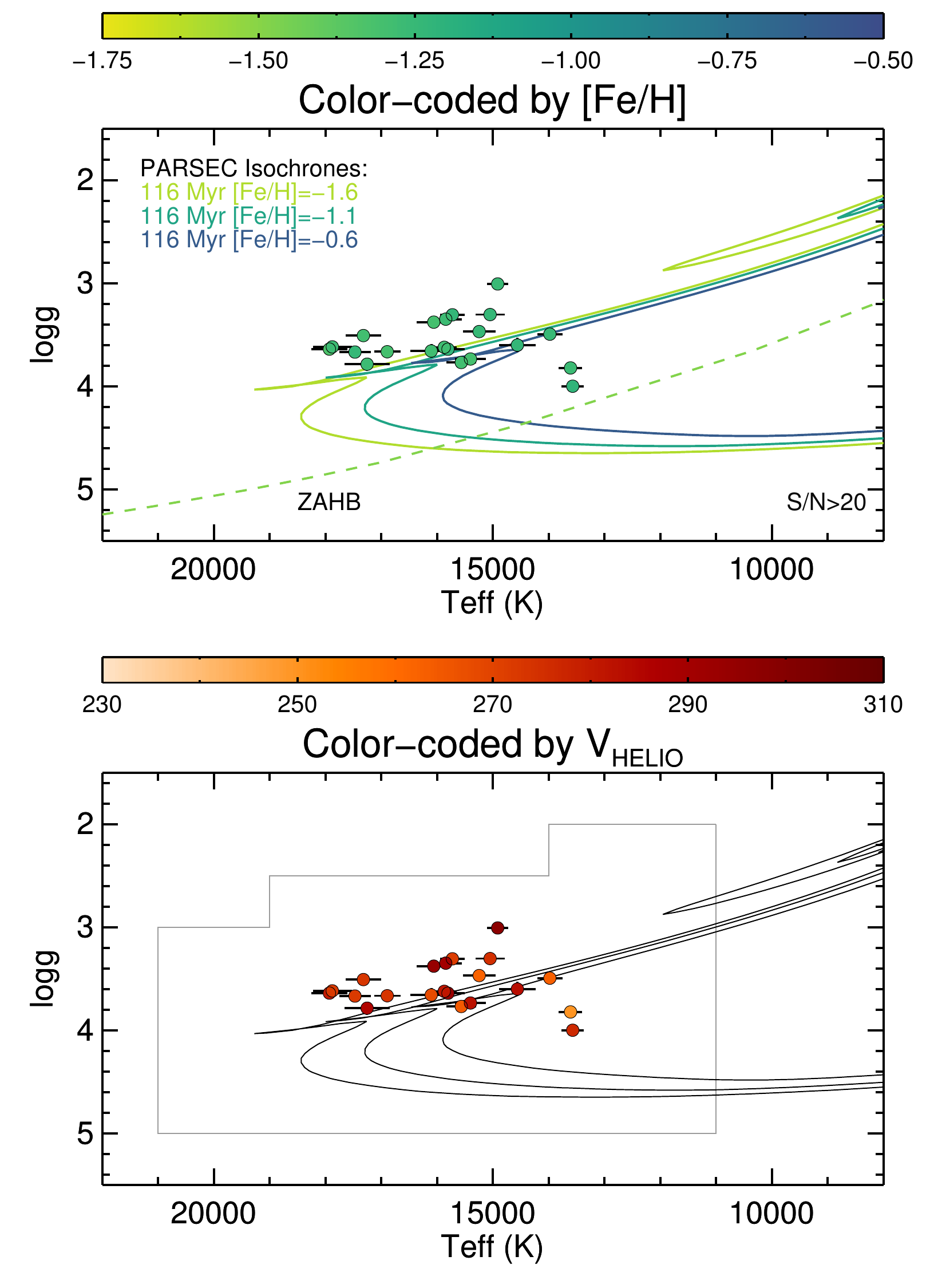}
\end{center}
\caption{Stellar parameters determined from \thecannon\ for the 21 \pwi stars with S/N$\ge$20 and compared to PARSEC isochrones.
The top panel shows \logg vs.\ \teff color-coded by [Fe/H] whereas the bottom shows the same points color-coded by \vhelioe.
As indicated by color coding in the top panel, the isochrones have an age of 116 Myr and a metallicity of [Fe/H] = $-$1.6, $-$1.1 and $-$0.6 dex, respectively. 
The [Fe/H]=$-$1.48 Zero-Age Horizontal Branch (ZAHB) from \citet{Dorman1993} is also shown as a dashed line.
The boundary of the synthetic spectral grid used in \thecannon\ analysis is shown in light gray in the bottom panel.}
\label{fig_tefflogg}
\end{figure}

\begin{figure}
\begin{center}
\includegraphics[width=1.0\hsize]{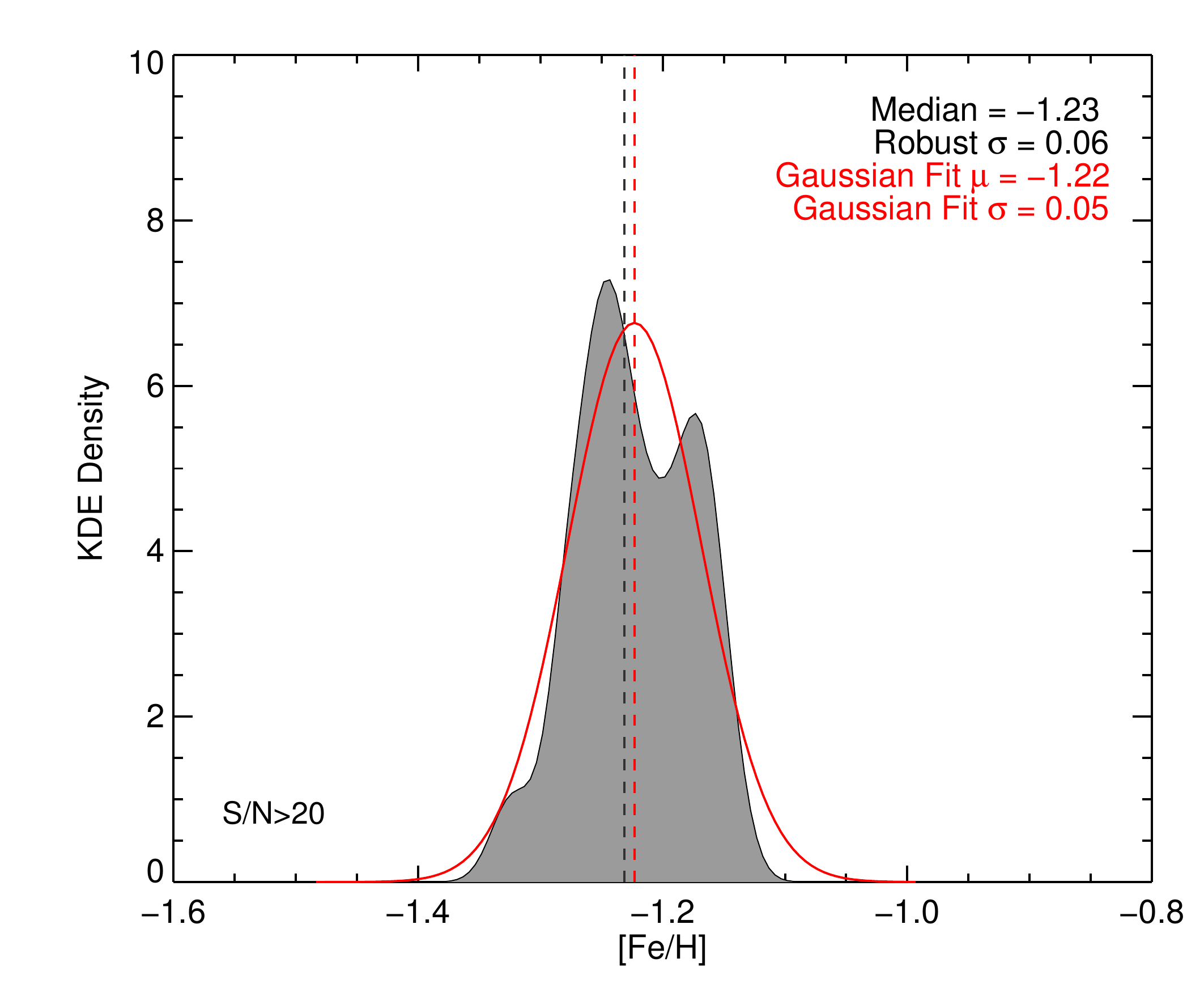}
\end{center}
\caption{Kernel Density Estimate (KDE) of the distribution of [Fe/H] measurements for 21 \pwi stars with S/N$\ge$20 and a bandwidth of 0.04 dex, which is smaller than the typical [Fe/H] uncertainty of $\sim$0.10 dex.  A single Gaussian model is fit to the histogram and has peak at $-$1.22 and $\sigma$ of 0.05 dex.  The median [Fe/H] is $-$1.23, the standard deviation is 0.06 dex.}
\label{fig_fehhist}
\end{figure}

\subsection{Stellar Parameters} \label{subsec:cannon}

We use \thecannon\footnote{\url{https://github.com/andycasey/AnniesLasso}} \citep{Casey2016,Ness2015} to determine stellar parameters (\teffe, \logge, [Fe/H]) for our 28 spectra.  The Cannon is a data-driven model for stellar spectra in which the stellar flux (at a given wavelength) is parameterized as a polynomial function of stellar parameters and abundances (i.e., ``labels'').  A Cannon ``model`` is trained on a set of spectra (observed or synthetic) with well-known labels (i.e., the ``training set'') and can then be used to determine the labels for other spectra.  An advantage of this method is that it exploits at the information available in the spectra and is also fully automated.
First, we construct a grid of 330 synthetic spectra (at a resolution of 1.5\AA) with the Synspec\footnote{\url{http://nova.astro.umd.edu/Synspec43/synspec.html}} \citep{Synspec,Hubeny2017} spectral synthesis software and IDL wrappers and auxiliary scripts (Allende Prieto, private communication).  Given an input atmospheric model, a linelist and stellar parameters and abundances, Synspec solves the radiative transfer equation over a specified wavelength range and resolution resulting in a synthetic spectrum.
We use the Kurucz LTE atmospheric models and include both atomic and molecular lines.
For the spectral grid, the \teff ranges from 11,000K to 21,000K in steps of 1,000K, the \logg ranges from 2.0 to 5.0 in steps of 0.5 dex, and the [Fe/H] ranges from $-$1.5 to $+$0.5 in steps of 0.25 dex (the boundary of the grid is shown in the bottom panel of \autoref{fig_tefflogg}).  
The synthetic spectra were interpolated onto the wavelength grid of the final MIKE spectra and normalized.  
For \thecannon\ model, we used a quadratic polynomial model to represent how the flux varies with the three stellar parameters at each wavelength.  
To provide a simple sanity check of the model generated by \thecannon, we determined the stellar parameters for the training set spectra using \thecannon\ model and compared these results to the training set input parameters.  We found no significant biases.
Specifically, the \teff had an offset of 0.5K and scatter of $\sigma$=250K, \logg had a mean offset of $-$0.003 dex and $\sigma$=0.06 dex, while [Fe/H] had no offset with a scatter of $\sigma$=0.12 dex.

\begin{figure}[t]
\begin{center}
\includegraphics[width=1.0\hsize, trim=0 4em 0 0]{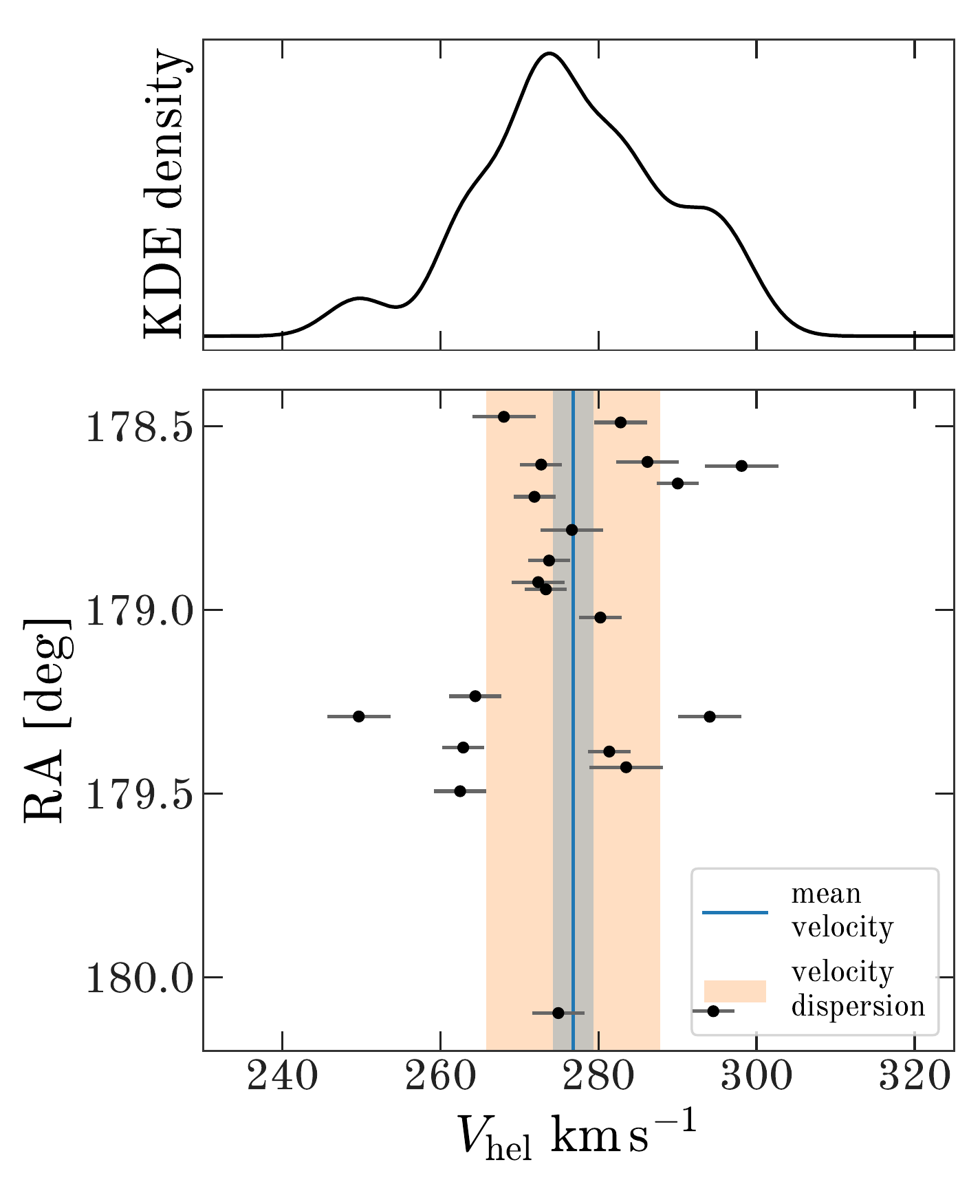}
\end{center}
\caption{\textit{Top panel}: Kernel density estimate of the distribution of RV measurements with a bandwidth $b = 3~\kmse$.
\textit{Bottom panel}: Individual RV measurements for \pwi stars with uncertainties (black markers and error bars) as a function of right ascension (RA).
The solid line (blue) shows the inferred mean RV of \pwie, and the shaded (blue) region shows the uncertainty on the mean RV, $276.7 \pm 2.5~\kmse$.
The larger shaded region (orange) shows the median posterior value of the velocity dispersion of \pwie, $\measvdisp \pm 2~\kmse$.
}
\label{fig:rv_inferred}
\end{figure}

We use the initial stellar parameters determined with \thecannon\ to refine the normalization of the MIKE spectra using a scaling factor determined by the ratio of the observed spectrum divided by the best-fit Cannon spectrum with heavy Gaussian-smoothing (FWHM = 38\AA). New \cannon\ solutions were then found for these ``renormalized'' observed spectra. Two example fits from \thecannon\ are shown in Figure~\ref{fig_spectra_model} in comparison to their best-fit \cannon\ model and its parameters.

We use the {\em emcee} \citep{emcee, Goodman:2010} MCMC sampler to determine the stellar parameters for each star and their uncertainties.
We run with 30 walkers for 1000 step and the first 200 are discarded as ``burn-in'' steps. \autoref{fig_cannoncorner} shows an example ``corner'' plot of the posterior distributions for PW1-18 and the median values with the blue lines.
Typical statistical uncertainties are $\sim$300K in \teffe, 0.06 dex in \logge, and 0.10 dex in [Fe/H].

\autoref{fig_tefflogg} shows the distribution of \logg vs. \teff color-coded by [Fe/H] and \vlsr in comparison to PARSEC isochrones. We note that one or two stars (including PW1-13) might be hot horizontal branch (HB) stars based on the best-fit \logg and \teff \citep[e.g.,][]{Zhang2019}. Confirming whether a star is either on the HB or MS evolutionary stage is beyond the scope of this paper due to the lack of He abundance information. However, we emphasize that inclusion of a few potential HB stars does not affect our dynamical analysis and main conclusions  because all the stars in our sample share consistent proper motions and radial velocities.

\autoref{fig_fehhist} shows the resulting [Fe/H] distribution of \pwi stars, which is peaked at [Fe/H] $\approx$ $-$1.23 and has a dispersion of 0.06 dex.  This spectroscopically-measured metallicity is in good agreement with our photometry-based metallicity measurement of [Fe/H]=$-$1.1 in \citetalias{PriceWhelan2018}. The small dispersion of the \pwi [Fe/H] values suggests that our uncertainties might be overestimated and that our true uncertainties are closer to $\sim$0.05 dex.  The derived stellar parameters and their uncertainties are also provided in Table 1.

\section{Results} \label{sec:results}


\begin{figure*}
\begin{center}
\includegraphics[width=1.0\hsize]{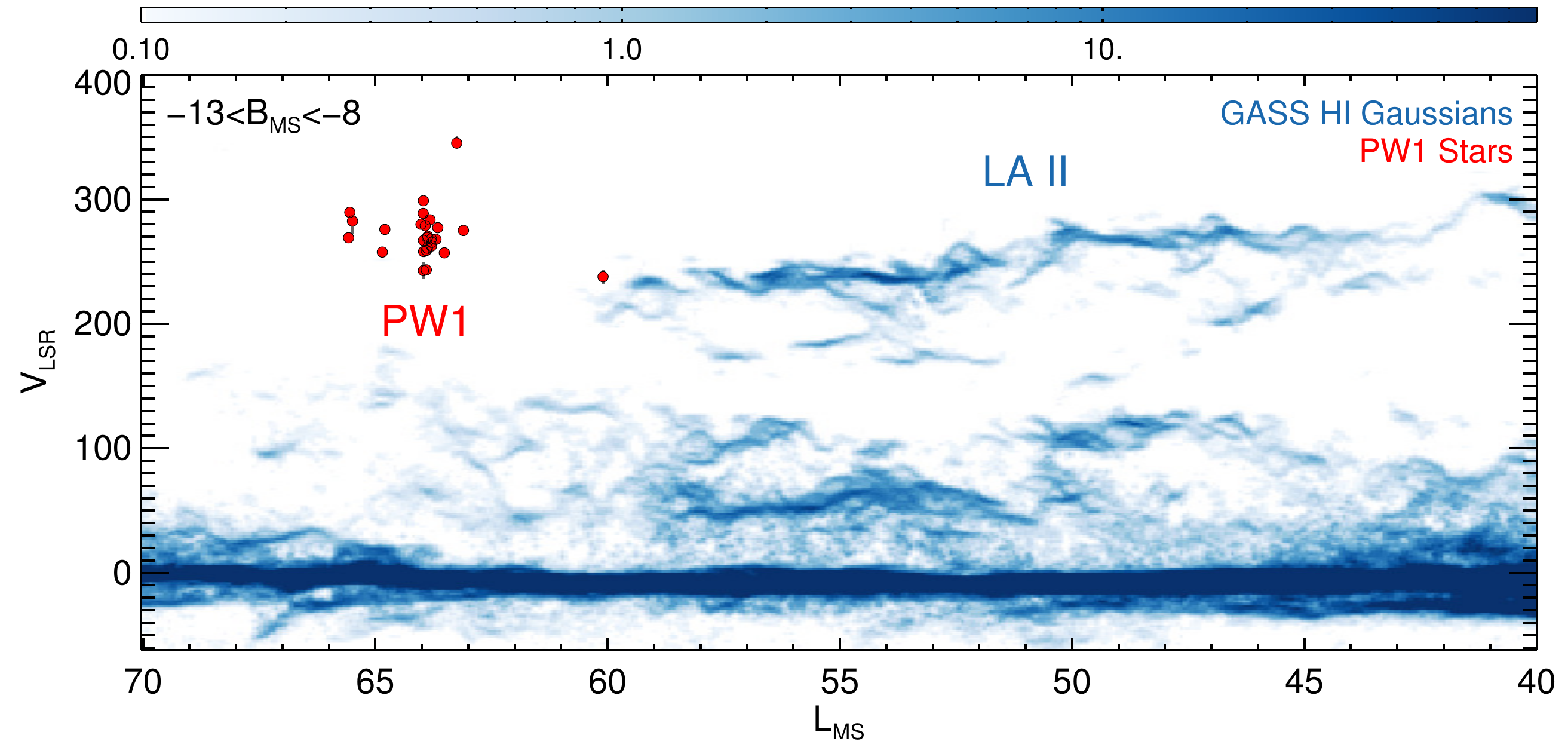}
\end{center}
\caption{Position--velocity diagram of the \pwi stars compared to the GASS \hi data \citep{McClure-Griffiths2009}.  
The background bluescale image is the integrated intensity of the GASS Gaussian centers summed along \bms in units of 10$^{18}$ atoms cm$^{-2}$ degree 
The tip of \laii has a velocity of $\sim$233 \kmse.  
Each \pwie star is plotted at its determined \vlsre as a red filled circle with its uncertainty. 
The stars have a mean velocity of $\sim$273.4 \kms, slightly higher than but similar to the gas in LA II.}
\label{fig_vlsrmlon}
\end{figure*}

\subsection{\pwi Velocity} \label{subsec:dynamics}

With proper motion and radial velocity information for individual stars in \pwie, along with a constraint on the mean distance to \pwi \citepalias{PriceWhelan2018}, we construct a simple model for its internal velocity structure to infer the true mean velocity and velocity dispersion from these projected quantities.
We use the $N=22$ (out of 28) stars with spectroscopic $\textrm{S}/\textrm{N} \geq 10$ for this analysis. 

We assume that the 3-D velocity of each $n$ star, $\vec{v}_n$, is drawn from a 3D Gaussian with mean $\vec{v}_0$ and a diagonal covariance matrix $\mat{C} = \sigma_v^2 \, \mathbb{I}_3$, where $\sigma_v$ is the (assumed isotropic) velocity dispersion, and $\mathbb{I}_3$ is the identity matrix.\footnote{We also tried allowing generic velocity anisotropies (non-diagonal $\mat{C}$) but found that the resulting matrix was consistent with being isotropic within the derived uncertainties.}
For each individual star, we only observe projected or astrometric quantities, like sky position $(\alpha, \delta)$, distance $d$, proper motions $(\mu_\alpha, \mu_\delta)$, and radial velocity $v_r$.
We assume that the sky position of each star is known with infinite precision, the observed distance is unresolved and given by the mean cluster distance and uncertainty \citepalias{PriceWhelan2018}, the proper motions are given by \textit{Gaia} with a 2D covariance matrix $\mat{C}_\mu$, and the radial velocities are measured in this work.
The (unobserved) true 3D velocity of each star, drawn from the model given above, is related to the observed astrometric quantities through a  projection matrix that depends on sky position (see, e.g., the appendix of \citealt{Oh2017}).
The full model therefore has $N + 3\,N + 3 + 1 = 92$ parameters: the true distance to each star, the true 3D velocity vector of each star, the mean velocity of \pwie, and the velocity dispersion.
This model is implemented in the \texttt{Stan} \citep{stan} probabilistic programming language, and we use the built-in No-U-Turn Hamiltonian Monte Carlo sampler \citep{NUTS} to generate posterior samples over all of the model parameters. 
We run the sampler for 4000 steps in total for 4 independent Markov Chains: 2000 burn-in and tuning steps (that are discarded), and then a further 2000 steps for each chain, from which we assess convergence by computing the effective sample size and Gelman-Rubin convergence statistic for each chain.

Given the posterior samples generated as described above, we measure a mean barycentric radial velocity for \pwi of $\measvr \pm 2.5~\kmse$ and mean proper motion of $(-0.52, 0.42) \pm (0.04, 0.03)~\masyr$.
Assuming a total solar velocity of $(11.1, 220 + 12.24, 7.25)~\kms$ \citep{Schoenrich2010}, this corresponds to a Galactocentric (Cartesian) velocity for \pwi of $v_{\textrm{galcen}} \approx (0.6, 2.2, 186)~\kmse$.
We find a velocity dispersion of $\sigma_v = \measvdisp \pm 2~\kmse$, which is consistent with a robust standard deviation computed from the RVs alone (\S \ref{subsec:rv}), indicating that the constraint on the velocity dispersion comes predominantly from the RV data in this work.
\autoref{fig:rv_inferred} shows the individual RV measurements of each observed \pwi star (black points) as a function of right ascension, along with the inferred mean velocity (blue) and velocity dispersion (orange).

In the sections to follow, we compare the inferred line-of-sight velocity of \pwi with \hi radio data (e.g., from GASS) where it is necessary to convert to the ``kinematic local standard of rest'' (LSRK), which historically uses the average of Solar neighborhood star velocities \citep{Delhaye1965,Gordon1976}\footnote{The solar motion is assumed to be 20.0 \kms towards (18h, +30\degr) at epoch 1900.0.}.
In this reference frame, the mean velocity is $\vlsre = \measvlsrk \pm 2.5~\kmse$.


\subsection{Leading Arm Origin} \label{subsec:laorigin}

Figure~\ref{fig_vlsrmlon} shows the position--velocity diagram of the \pwi stars and the \hi gas \citep[GASS;][]{McClure-Griffiths2009} using the Gaussian decomposition techniques described in \citet{Nidever2008}. The mean \vlsr velocity of \measvlsrk~\kms of the \pwi stars is very similar to the velocity of the \laii gas of 233 \kmse. 
Absorption-line studies of the LA HI find a value of [Fe/H]$\sim$-1 dex \citep{Lu1998,Fox2018,Richter2018}, which is similar to the mean metallicity of \pwi stars studied here of [Fe/H]$\sim$--1.23 dex (see \autoref{fig_fehhist}). 
Therefore, we conclude that the \pwi stars are physically associated with the \laii gas based on their similar RVs and metallicities. 

\begin{figure}[t] 
\begin{center}
\includegraphics[width=1.0\hsize,angle=0]{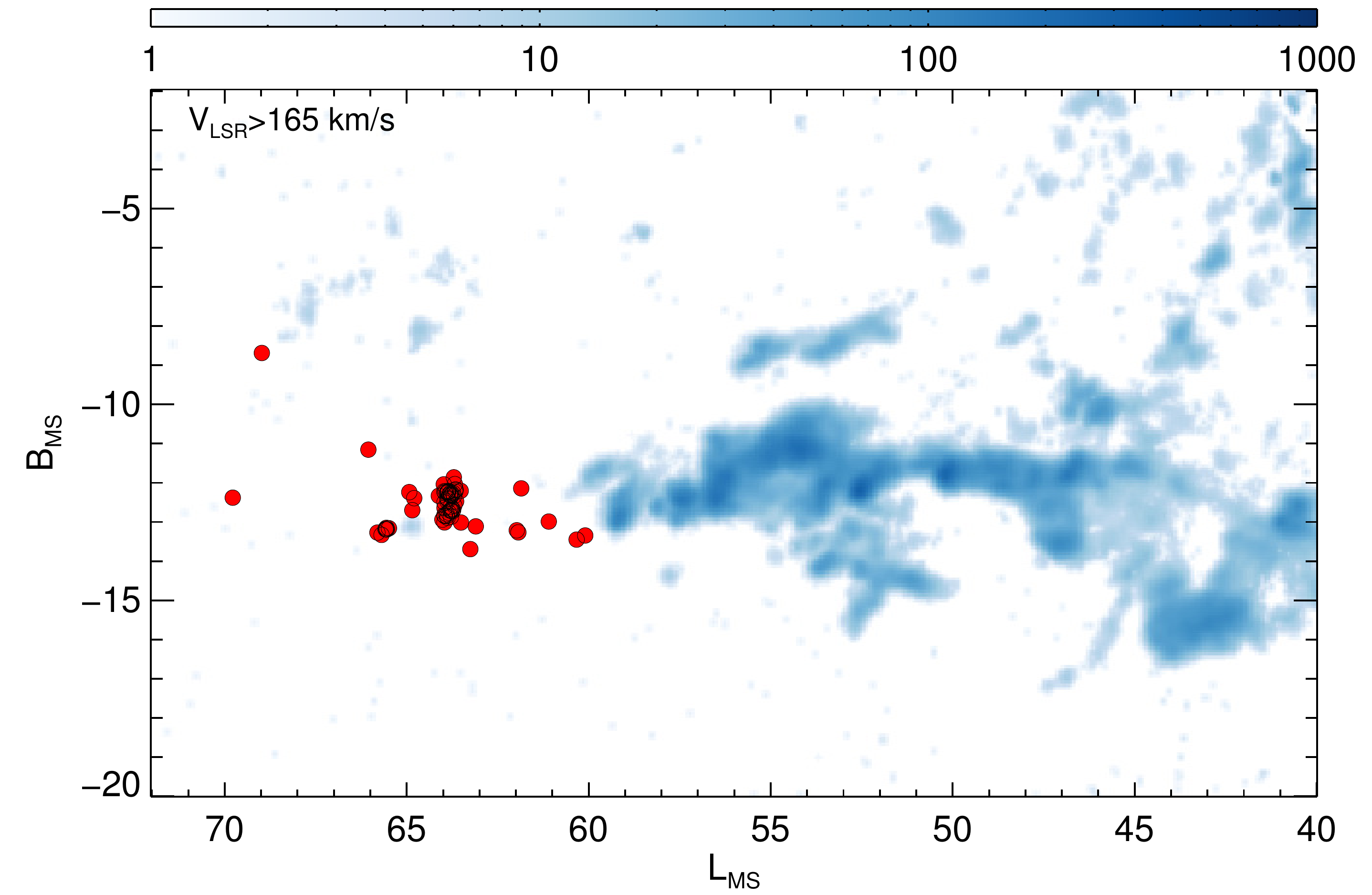} \\
\includegraphics[width=1.0\hsize,angle=0]{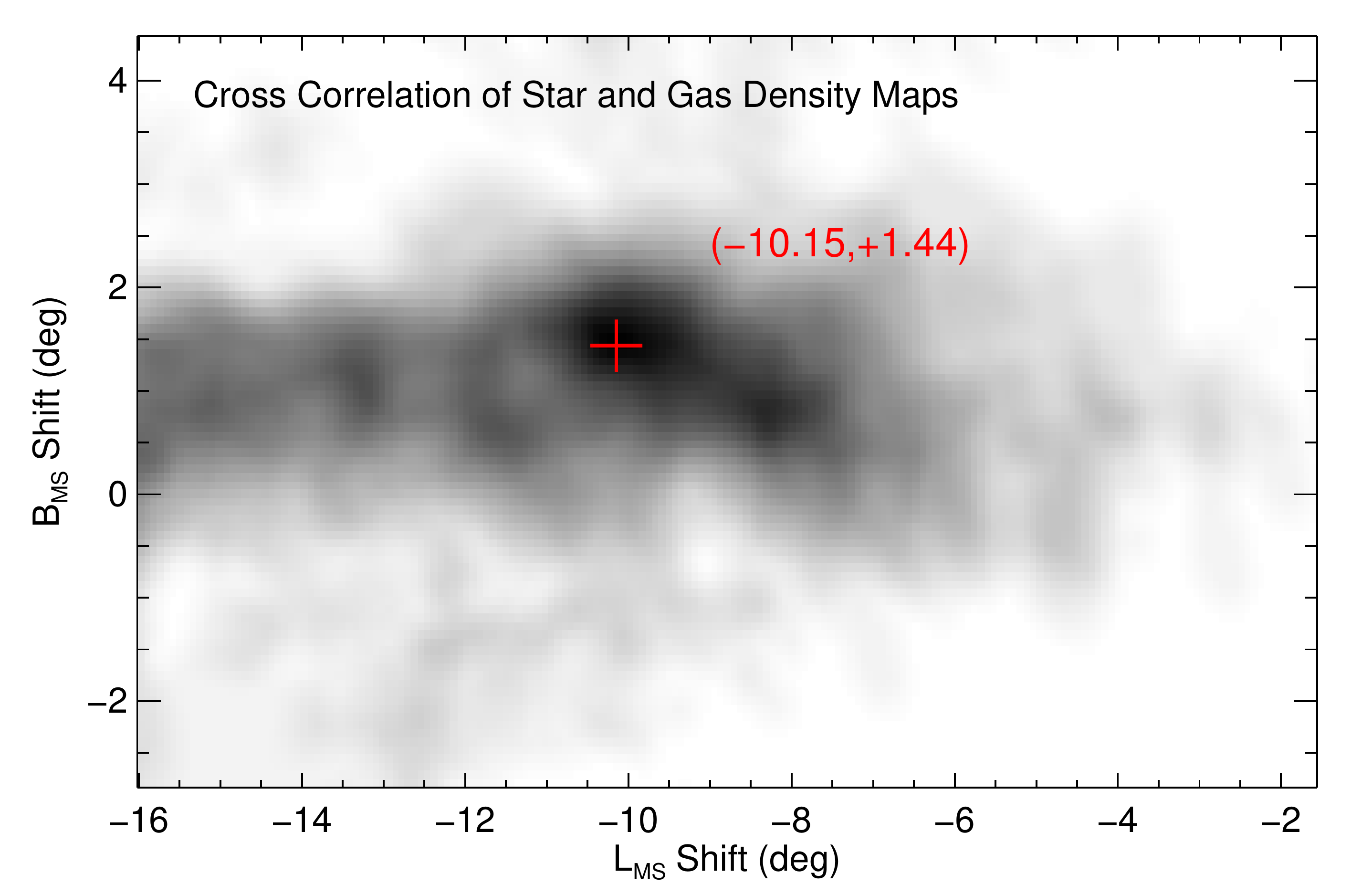} \\
\includegraphics[width=1.0\hsize,angle=0]{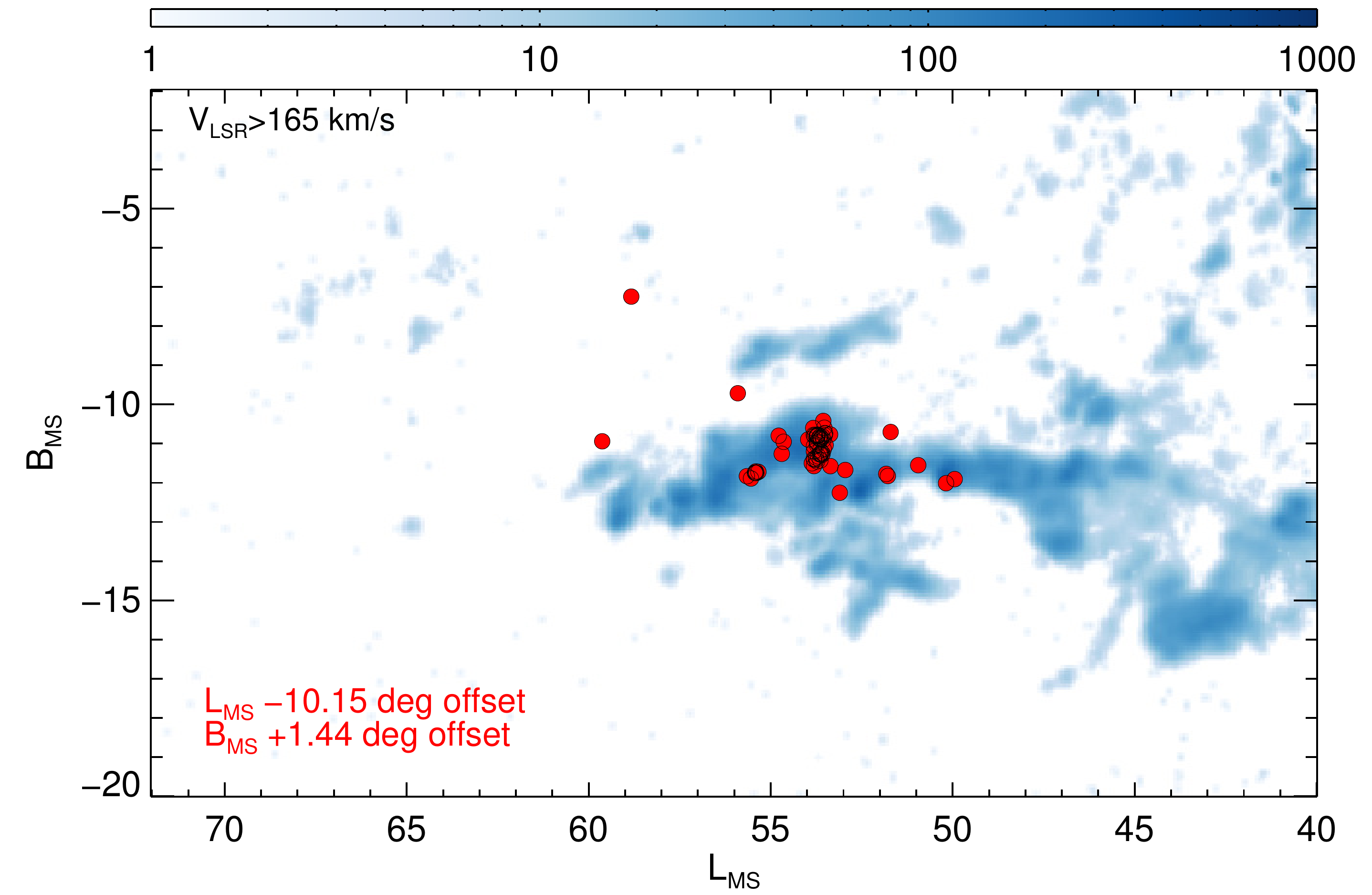}
\end{center}
\caption{ The spatial distribution of the \hi gas and 71 \pwi candidate stars based on photometry and Gaia DR2 proper motions. 
(top) The gas and \pwi stars at their current (\lmse,\bmse) positions. 
(middle) The 2D cross-correlation map of the star and gas density maps with the shift of ($-$10.15\degr,$+$1.44\degr) in (\lmse,\bmse).
(Bottom) \pwie shifted by the offset determined in the cross-correlation.}
\label{fig_shiftmap}
\end{figure} 

\begin{figure} 
\begin{center}
\includegraphics[width=1.0\hsize,angle=0]{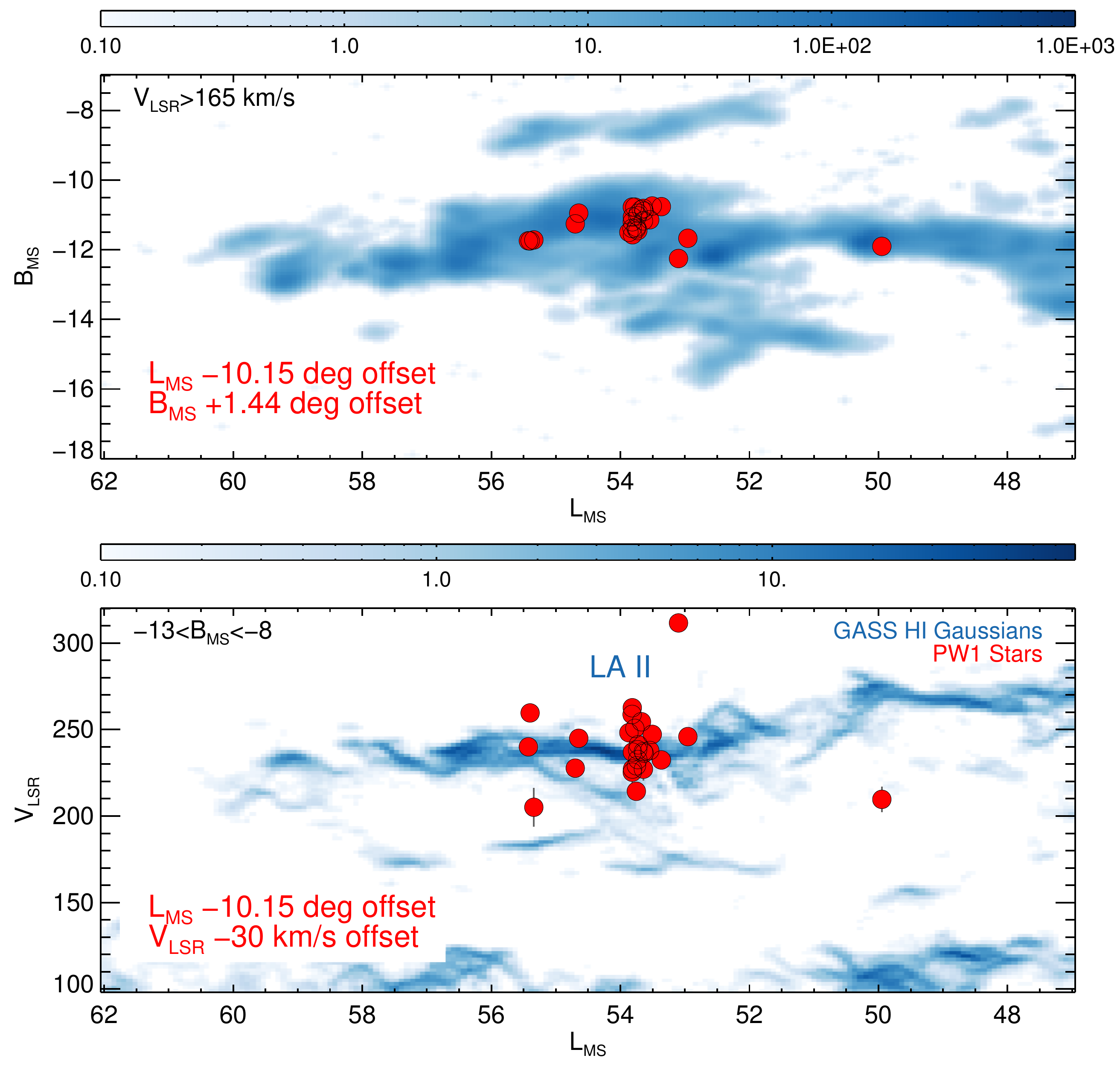}
\end{center}
\caption{Spatial and Position--velocity diagram for the 28 \pwi stars with radial velocities and \laii gas.
The top panel shows the positions of the \pwi stars (red filled circles) and the \laii gas after the cross-correlation offset has been applied (see \autoref{fig_shiftmap}). 
The bottom panel shows the same spatially-offset stars in position-velocity space. 
A \vlsr offset of 30 \kms has been applied such that the stars align with the densest portions of the \laii gas. 
}
\label{fig_offset_2panel}
\end{figure} 

\subsection{Spatial Offset between Stars and Gas} \label{subsec:spatialoffset}

The current location of \pwi is close to \laii on the sky, but there is little gas in its immediate vicinity in the position-velocity diagram (\autoref{fig_vlsrmlon}). The top panel of \autoref{fig_shiftmap} shows the spatial offset between \pwi and the \laii gas in the MS coordinate system (\lmse,\bmse). 
The gas from which \pwi originated should experience ram pressure from the MW hot halo gas \citep[e.g.,][]{Mastropietro2005}.  Because the stars will not feel this force, the gas and stars will decouple over time with the net effect that the gas trails the stars in their orbit. 
With this in mind, it is not surprising that there is little gas in the immediate vicinity of \pwie. Thus, the spatial offset does not rule out \laii as the origin gas for \pwie. 

Quantitatively, the magnitude of the spatial decoupling between the stars and gas is determined by a 2D cross-correlation of the star and gas maps (\autoref{fig_shiftmap}, top). 
The individual maps were masked such that only the highest density regions were used.  
The result of the cross-correlation is in the middle panel of \autoref{fig_shiftmap}.  
The peak of the cross-correlation is at an offset of ($-$10.15\degr,$+$1.44\degr) in (\lmse,\bmse), which is indicated as a red cross in the middle panel of \autoref{fig_shiftmap}.
The bottom panel of \autoref{fig_shiftmap} shows that shifted \pwi positions are well aligned with the high-density peaks of the \hi gas. 
These shift positions are more representative of the original relative positioning of the gas forming \pwi within the \laii complex.

\autoref{fig_offset_2panel} shows a slightly different visualization of this process. 
The top panel of \autoref{fig_offset_2panel} zooms into the (\lmse,\bmse) region around the position-shifted stars with \vlsr measurements. 
Ram pressure will also act to modify the velocities of the gas it acts on, and thus an exact match between the velocities of the birth-gas and resulting stars is not expected.
The bottom panel of \autoref{fig_offset_2panel} is the resulting velocity-position diagram; a 30 \kms velocity offset is required to have the \pwi stars align with the densest portions of the gas in position-velocity space. 
We will use this offset in the next section to place a constraint on the halo gas driving the ram pressure effects.

\subsection{Ram Pressure} \label{subsec:rampressure}

If the spatial and velocity offsets between this gas and \pwi are indeed due to a roughly constant ram pressure, then the measured position-velocity offsets become a means of inferring the properties of the gas performing the ram pressure. 
In particular, we place a constraint on the average density of the ``hot'' MW-halo gas, a medium that remains difficult to characterize directly \citep[for an overview see][]{putman2012}.

At a distance of 29 kpc, the angular offset of 10.15\dgr corresponds to a tangential distance of $\sim$5.14 kpc.  Over 116 Myr that corresponds to a mean ``drift" velocity of $\sim$43.3 \kms or a deceleration of $\sim$0.747 \kmsmyre.  
The total change in the tangential velocity of the gas over this period is $\sim$86.6 \kmse.
The \vlsr of \laii ``tip" is lower than that of \pwi by $\sim$30 \kmse. 
The smaller \vlsr shift would indicate that a significant fraction of the deceleration occurred in the tangential direction.


If we assume that ram pressure is the main factor in causing the gas and stars to drift apart and that this is dominated by a roughly constant ram pressure from the
MW hot halo, then we can estimate the density of the MW hot halo gas as follows.
The ram pressure experienced by the \pwi ``birth cloud" (BC) in \laii is defined as,
 \begin{equation} \label{eq_ram} 
   P = \rho_{\rm MW} ~ v_{\rm BC}^2, 
 \end{equation}
where $\rho_{\rm MW}$ is the density of the MW halo gas and $v_{\rm BC}$ is the birth cloud velocity through the medium.
If $l_{\rm BC}$ is the approximate diameter of the birth cloud, then the acceleration that it experiences ($a_{\rm BC}$) is,
 \begin{equation} \label{eq_accel} 
  a_{\rm BC} 
  \approx  \frac{\rho_{\rm MW} ~ v_{\rm BC}^2 ~ l_{\rm BC}^2}{\rho_{\rm BC} ~ l_{\rm BC}^3}  
  \approx \frac{\rho_{\rm MW} ~ v_{\rm BC}^2}{\rho_{\rm BC} ~ l_{\rm BC}}.
 \end{equation}  
Solving \autoref{eq_accel} for the ratio of gas density between the MW and the birth cloud gives 
 \begin{equation} \label{eq_densratio} 
   \frac{\rho_{\rm MW}}{\rho_{\rm BC}} \approx \frac{a_{\rm BC} ~ l_{\rm BC}}{v_{\rm BC}^2}.
 \end{equation} 
If we assume that the birth cloud was undergoing a roughly constant deceleration due to ram pressure since forming \pwie, then we can estimate this deceleration from the age of \pwi ($\Delta{\rm t}$ = 116 Myr) and the spatial offset ($\Delta{\rm x}$ = 10.15\dgr = 5.14 kpc), as
 \begin{equation} 
   a_{\rm BC} \approx \frac{2 \Delta{\rm x}}{\Delta{\rm t}^2}.
 \end{equation}
This results in $a_{\rm BC} \approx$ 0.747 \kmsmyre.
The angular width (full-width at half-maximum in \bmse) of the \laii \pwi birth cloud from GASS is 0.75\dgr or 0.38 kpc.  Using this as an approximation of the diameter of the birth cloud ($l_{\rm BC}$) and $v_{\rm BC}$ $\approx$ 235 \kmse, we obtain $\rho_{\rm MW}/\rho_{\rm BC}$ $\approx$ 0.00505 or that the MW gas density is $\sim$0.5\% of that of the \pwi birth cloud. 

We measure the average current column density of the \pwi birth cloud from the \hi GASS data as \nhi $\approx 1.5 \times 10^{20}$ atoms cm$^{-2}$.
Assuming a distance of 29 kpc and a width of $\sim$0.38 kpc gives a number density of $n_{\rm BC} \approx$ 0.128 atoms cm$^{-3}$. 


Finally, we derive the number density of the hot MW halo as $n_{\rm MW} \approx$ 6.46 $\times 10^{-4}$ atoms cm$^{-3}$.  
This rough estimate of the hot MW halo gas density is an order of magnitude higher than that predicted by the \citet{Miller2013} model which gives $\sim$4$\times$10$^{-5}$ atoms cm$^{-3}$ at this location.  
However, our estimate here is too simplistic for a number of reasons: (i) we are seeing an integrated effect over the orbit of \pwie, which has passed through higher densities closer to the midplane, and (ii) we consider only a single medium -- the hot halo -- completely ignoring the impact of the midplane and outer gas disk itself.
That this simple estimate is within an order-of-magnitude of state-of-the art measurements motivates the more complex and nuanced simulations that are presented in the next sub-section.  

\subsection{Ram pressure orbit analysis}\label{subsec:orbit}

The offsets in sky position and velocity between the \pwi stars and the \laii gas suggest that the gas was likely subject to a drag force or dissipated orbital energy: The offset in position is primarily along the direction of motion, and the gas velocity is slightly lower than that of \pwie.
Moreover, the order-of-magnitude computations presented in the previous sub-section suggest that ram pressure from the hot halo produce effects on par with what is seen, albeit more complex interactions of the mid-plane are ignored.
We therefore perform a set of orbit integrations to see if the observed position-velocity offsets can plausibly be explained by gas drag from the MW halo when the full orbital motion and influence of the midplane are included.

We use the MW mass model implemented in \texttt{gala} \citep{Bovy2015galpy, PriceWhelan2017} as the background gravitational potential, and include a time-dependent mass component to represent the LMC. 
In detail, we represent the mass distribution of the LMC using a Hernquist profile \citep{Hernquist1990}, which generally follows the the approach used in \citet{Erkal2019}, except that we use a total mass of $M_{\rm LMC} = 2.5\times 10^{11}~\msune$ \citep{Laporte2018b}.
We compute the position of the LMC at any given time by integrating the orbit of the LMC center-of-mass backwards from its present day phase-space position (using initial conditions from \citealt{Patel2017}).
We neglect any back-reaction or response of the MW to the presence of the LMC.
We use this time-dependent mass model along with the measured position and velocity of \pwi to numerically integrate the orbit of the \pwi backwards in time from present day ($t_2=0$) to its birth time ($t_1 = -\tau = -116~{\rm Myr}$).
We use the Dormand-Prince 8th-order Runge-Kutta scheme \citep{Dormand1980} to numerically integrate these orbits.

Once the phase-space coordinates of \pwi at its birth time are estimated, we then integrate forward in time, now including the effects of gas drag and momentum coupling between the MW disk and the \laii gas as it crosses the Galactic midplane. 
We compute the drag acceleration on the gas as 
\begin{align}
    \bs{a}_{\rm drag} &= n_{\rm MW}(\bs{x}) \, |\bs{v}|^2 \, \Sigma_{\rm LA}^{-1} \, \times \, -\frac{\bs{v}}{|\bs{v}|}
\end{align}
where $n_{\rm MW}(\bs{x})$ is the number density of MW gas at position $\bs{x}$, $\bs{v}$ is the orbital velocity, and $\Sigma_{\rm LA}$ is the surface density of the LA gas \citep[following, e.g.,][]{Vollmer2001}.
For the gas density, we use the gaseous halo model from \citet{Miller2013} and the disk model from \citet{Kalberla2009} with a Gaussian density profile in the height above the midplane $z$ such that
\begin{align}
    n_{\rm MW}(\bs{x}) &= n_{\rm halo}(r) + n_{\rm disk}(R, z)\\
    n_{\rm halo}(r) &= n_{0, h} \, \left[1 + \left(\frac{r}{r_c}\right)^2\right]^{-3\beta/2} \\
    n_{\rm disk}(R, z) &= n_{0, d} \, e^{-\frac{R-R_\odot}{R_n}} \, e^{-\frac{1}{2} \, \frac{z^2}{\sigma_z(R)^2}}\\
    \sigma_z(R) &= 0.85 \, h_0 \, e^{\frac{R-R_\odot}{R_0}}
\end{align}
where $r$ is the spherical radius, $R$ is the cylindrical radius, $R_\odot = 8.1~\kpc$ is the solar Galactocentric radius, and all parameter values are taken from \citet{Kalberla2009}.
We assume that the surface density of the \pwi birth cloud in \laii starts with $\Sigma_0 \approx 50~\msune~\textrm{pc}^{-2}$ \citep[comparable to other LMC molecular clouds][]{Wong2011} and ends (at present day) with a surface density equal to the measured column density of the \laii region, $\Sigma_f \approx 0.56~\msune~\textrm{pc}^{-2}$ such that
\begin{equation}
    \Sigma_{\rm LA}(t) = \exp\left[\frac{(\ln\Sigma_f - \ln\Sigma_0)}{\tau} \, t + \ln\Sigma_f\right]
\end{equation}
where again $\tau$ is the age of \pwie, $t_1 = -\tau$, and $t_2=0$.

As mentioned above, we also allow for momentum coupling between the MW disk and the \laii gas as it passes through the Galactic midplane.
To take this momentum coupling into account, we add an additional acceleration to the orbit integration, $\bs{a}_{\rm coupling}$, defined to point in the direction of Galactic rotation such that
\begin{equation}
    \bs{a}_{\rm coupling} = \alpha \, \hat{\bs{v}}_{rot}(\bs{x}) \, e^{-\frac{1}{2} \, \frac{z^2}{\sigma_z(R)}}
\end{equation}
where $\alpha$ is a free parameter that determines the magnitude of the coupling, $\hat{\bs{v}}_{rot}(\bs{x})$ is a unit vector that points in the direction of Galactic rotation at the position $\bs{x}$, and the Gaussian in height, $z$, makes this operate only when the gas orbit is close to the Galactic plane.

We next allow the scale densities, $n_{0, h}$ and $n_{0, d}$, and the momentum coupling parameter, $\alpha$, to vary and fit for the values of these parameters that best reproduce the observed position and velocity offsets between \pwi and the \laii gas.
We define a fiducial point in \lmse, \bmse, \vhelio to set the present-day location of the densest \laii gas (see Section~\ref{subsec:spatialoffset}) that could have plausibly formed \pwie:
\begin{align}
    \lmse &= 53.7^\circ\\
    \bmse &= -11.1^\circ\\
    \vhelio &= 239~\kmse \quad .
\end{align}
We construct a likelihood function using the above orbit integration scheme and evaluate the likelihood of the present-day (i.e. final) orbit phase-space coordinates relative to the fiducial point defined previously.
We assume a Gaussian tolerance of $\sigma_{\rm LB} = 0.5^\circ$ for \lms and \bmse, and $\sigma_v = 1~\kmse$ for \vhelio and evaluate the likelihood as
\begin{align*}3
    p(\lmse, \bmse, \vhelioe &\given n_{0, h}, n_{0, d}, \alpha) = \\
    &\mathcal{N}(\textrm{L}_{\rm orbit} \given \lmse, \sigma^2_{\rm LB})\\
    \times &\mathcal{N}(\textrm{B}_{\rm orbit} \given \bmse, \sigma^2_{\rm LB})\\
    \times &\mathcal{N}(V_{\rm orbit} \given \vhelioe, \sigma^2_v)\numberthis
\end{align*}
where $\mathcal{N}(\cdot \given \mu, \sigma^2)$ represents the normal distribution with mean $\mu$ and variance $\sigma^2$.
In practice, we implement this function (programmatically) over the log values of the three parameters (because they must be positive), but we assume uniform priors in the parameter values over the domain $(a,b) = (e^{-30}, e^{5})$ for each parameter such that the prior distribution is
\begin{equation}
    p(n_{0, h}, n_{0, d}, \alpha) = \mathcal{U}(n_{0, h} \given a, b) \, \mathcal{U}(n_{0, d} \given a, b) \, \mathcal{U}(\alpha \given a, b)
\end{equation}
where $\mathcal{U}(\cdot \given a, b)$ is the uniform distribution defined over the domain $(a, b)$.

\begin{figure}[t]
\begin{center}
\includegraphics[width=1.0\hsize]{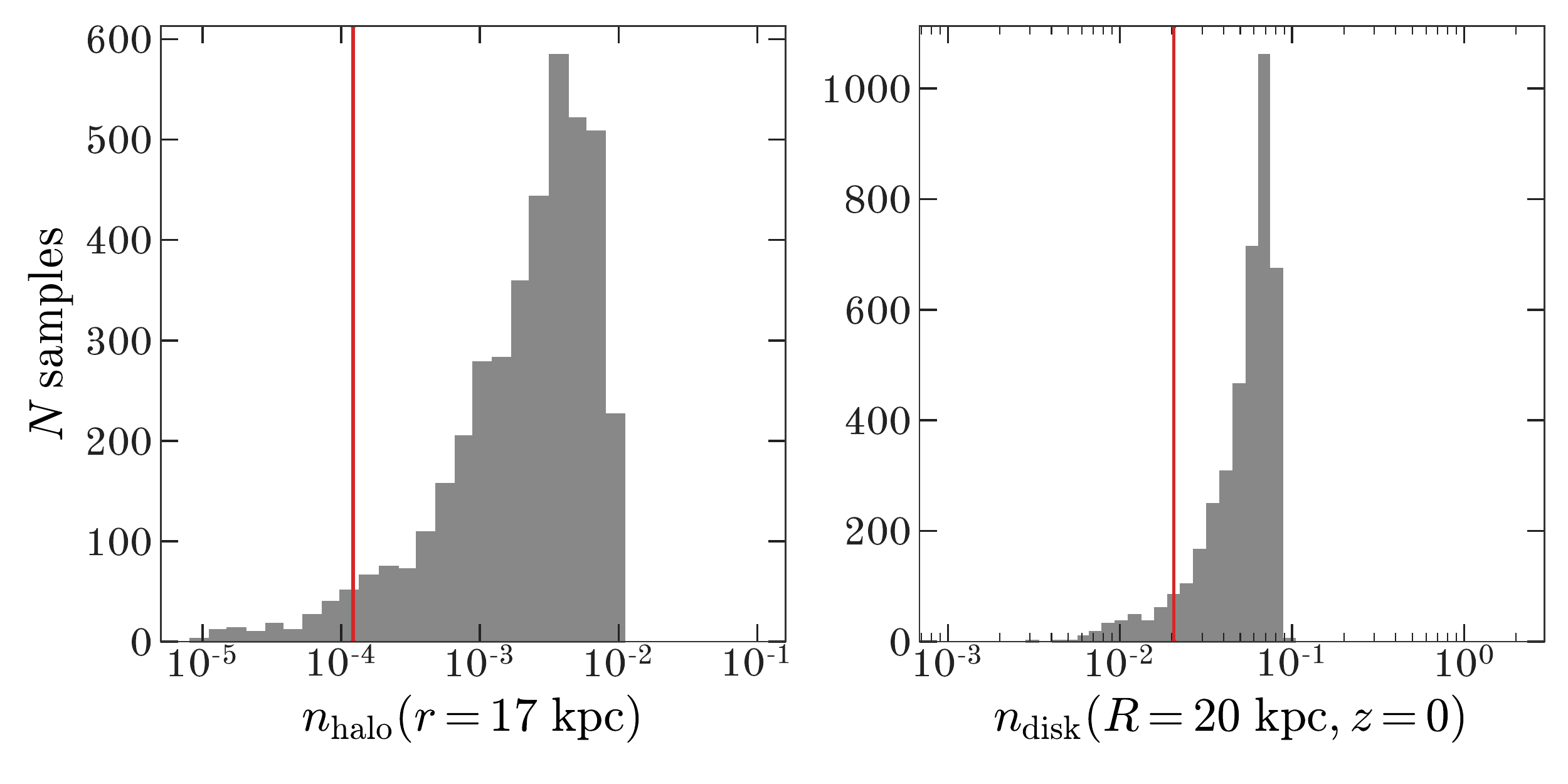}
\end{center}
\caption{Posterior samples from the ram pressure orbit analysis, transformed to values of the inferred halo gas density at $r=17~\kpc$ $n_{\rm halo}(17)$ (left panel) and disk gas density (right panel) at the midplane at $(R, z) = (20, 0)~\kpc$, $n_{\rm disk}(20, 0)$.
Vertical red lines show the values from the default Milky Way model from \citet{Miller2013} and \citet{Kalberla2009}.
}
\label{fig:diskhalodens}
\end{figure}

We first optimize the log-posterior, 
\begin{align*}
    \ln p(n_{0, h}, n_{0, d}, \alpha &\given \lmse, \bmse, \vhelioe) \propto \\
    &\ln p(\lmse, \bmse, \vhelioe \given n_{0, h}, n_{0, d}, \alpha) \\ 
    + &\ln p(n_{0, h}, n_{0, d}, \alpha) 
\end{align*}
using the BFGS algorithm \citep[implemented in \texttt{scipy},][]{bfgs, scipy} and then we use these optimal parameter values as initial conditions to run a Markov Chain Monte Carlo (MCMC) sampling of the posterior probability distribution (pdf) of our parameters.
We use an affine-invariant, ensemble MCMC sampler \citep[\texttt{emcee};][]{emcee, Goodman:2010}; we run with 64 walkers for 512 ``burn-in'' steps (that are discarded) and then run for an additional 1024 steps. 
We thin the resulting chains by taking every 4th step, and combine the parameter samplings from all thinned chains. 
Figure~\ref{fig:diskhalodens} shows histograms of posterior samples transformed into values of the halo gas density evaluated at the orbital pericenter, $n_{\rm halo}(17~\kpc)$, and the disk gas density at the midplane at $n_{\rm disk}(20~\kpc, 0)$.
The posterior values of the coupling coefficient, $\alpha$, were all $< e^{-12}$ and thus consistent with zero.

The best-fit parameters require a somewhat larger halo and disk gas densities than the fiducial MW gas density models from \citet{Miller2013} and \citet{Kalberla2009}.
In detail, we find
\begin{align}
    n_{\rm halo}(17~\kpc) &= 2.7_{-2.0}^{+3.4} \times 10^{-3}~{\rm atoms}~{\rm cm}^{-3}\\
    n_{\rm disk}(20~\kpc, 0) &= 6.0_{-2.0}^{+1.5} \times 10^{-2}~{\rm atoms}~{\rm cm}^{-3}
\end{align}
as compared to the fiducial values $n_{\rm halo, M13}(17~\kpc) = 1.2 \times 10^{-4}~{\rm atoms}~{\rm cm}^{-3}$ and $n_{\rm disk, K09}(20~\kpc, 0) = 2 \times 10^{-2}~{\rm atoms}~{\rm cm}^{-3}$.
However, the goal of this analysis is only to illustrate that the observed offsets could plausibly be described by ram pressure, and that the inferred MW halo and disk gas densities needed to explain the magnitude of the ram pressure drag are reasonable.
In doing this, we neglect more complex density evolution of the LA gas, assume that the \laii gas, at least around the \pwi birthplace, acts like a cloud (rather than a dissolving and morphologically-varying gas filament), and assume that no supernovae (SNe) have impacted the orbital energy of the gas.
Still, this result motivates more detailed simulation of the interaction between the LA and the MW.

\begin{figure}[t]
\begin{center}
\includegraphics[width=1.0\hsize]{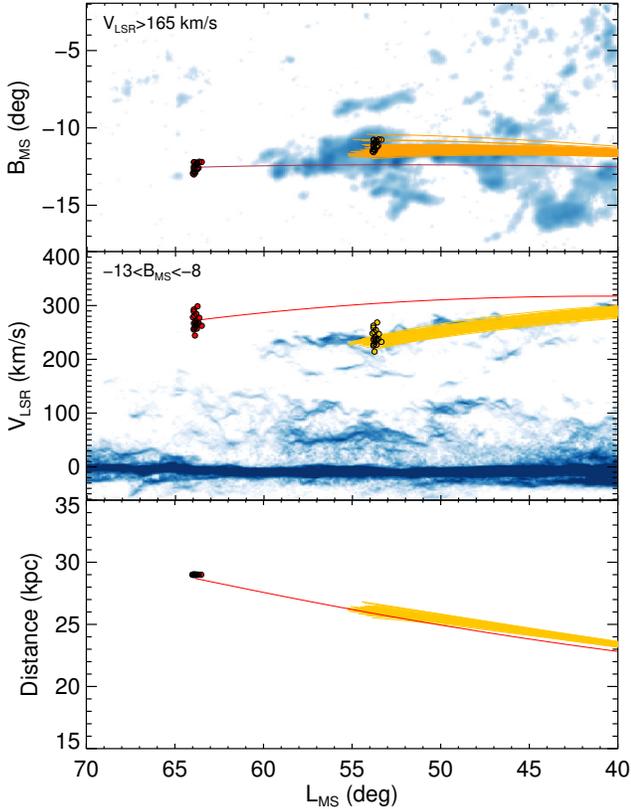}
\end{center}
\caption{The orbits of both the mean \pwi stars and the originating gas that experienced ram pressure. (Top) \bms vs.\ \lmse, (middle) \vlsr vs.\ \lmse, and (bottom) distance vs.\ \lmse.  The red line is the \pwi orbit with no ram pressure, while the orange lines show 64 orbits of the \pwi birth cloud drawn from the MCMC posterior distributions of the parameters.}
\label{fig_rampressureorbit}
\end{figure}

\section{Discussion} \label{sec:discussion}
The distance, radial velocity, metallicity, and orbit suggest that not only is \pwi associated with \laiie, but that it is also associated with the MCs and MS.  
The association of \pwi with \laii permits a more nuanced view of the LA than has previously been feasible. 
Not only does \pwi provide a distance measurement to the LA, but it also constrains its chemical, orbital, and dynamical properties, such as how it is affected by ram pressure from the MW hot halo.
At the same time, affiliating \pwi with the LA gas also explains some of its properties, as discussed below.

\subsection{The Spatial Morphology of \pwie}
One of the mysteries of \pwi is its unusual spatial shape and elongated distribution.  
However, the spatial correlation of the \hi \laii gas and the \pwi stars suggests a natural explanation for this. 
The \pwi stars do not represent one single cluster that has disrupted but rather is likely the outcome of multiple star formation events associated with high-density \hi clumps in \laiie. This is a common feature observed in jellyfish galaxies experiencing ram pressure in galaxy clusters.
Therefore, it might be more appropriate to call \pwi a star formation ``complex" or association rather than a star cluster in the traditional sense.

\subsection{Origin of the Leading Arm} \label{subsec:mcassocation}
The mean metallicity of \pwi of $\approx -1.23$ is similar to the measured metallicity of the LA ([O/H]$\approx$ $-$1.16; \citealt{Fox2018}), the Magellanic Bridge \citep[$\approx$ $-$1.0;][]{Lehner2008} and the trailing MS \citep[$\approx$ $-$1.2;][]{Fox2013a}.  
There is a large range in the measured metallicities of MW \hi high-velocity clouds \citep[HVCs;][]{Wakker2001}, and, therefore, the similarity of the metallicities in these distinct systems supports the notion that the LA, MB, and MS all share a common origin.
These metallicities are also consistent with the metallicity of the SMC $\sim$2~Gyr ago.
Therefore, all of these gas structures associated with the MCs likely originated mainly from the SMC from the same tidal event about 2~Gyr ago. 

Despite the similar metallicities among the gas structures, the origin of the MS and LA are still debated. This is mainly because (1) the observational data is far from complete -- e.g., the metallicity measurements are limited to small number of sightlines, and (2) there are some observed features that cannot be easily explained by the sole SMC origin \citep[e.g.,][]{Fox2013b}. 
One of the recent theoretical studies \citet{Pardy2018} argued that both the LMC and SMC contributed to create the LA and MS gas features. However, another recent MS simulation work by \citet{Tepper-Garcia2018} suggested that the LA gas does not originate in the MCs because the ram pressure from the MW hot halo gas would prevent the gas from reaching its present position. 
If \pwi is indeed affiliated with \laiie, as we suggest here, then there is now observational evidence that the impact of ram pressure is overestimated in \citet{Tepper-Garcia2018}.
The key discriminant is the assumed $\rho_{\rm MW}$ in \citet{Tepper-Garcia2018} versus what we infer from our scenario for \pwie.

\subsection{No Natal Gas Disruption?}
The conditions for triggering star formation in the LA is not well understood. Based on the fact that \pwi is the only known stellar component to date that is likely associated with \laiie,
the birth cloud of \pwi must have satisfied very special star formation conditions in the LA while passing through the Galactic midplane. Aside from the unknown star formation conditions of \pwie, there is another mystery: How has the morphology of the \pwi birth cloud remained mostly intact? Our analysis in \S \ref{subsec:spatialoffset} shows that the present-day spatial distribution of the associated \hi gas resembles that of the \pwi stars across $\sim$2.5 kpc.

A gas cloud that forms a young star cluster is disrupted when the first SN occurs. 
The SN-explosions effectively act to distort the original spatial correlations between the gas and stars by injecting radiative and mechanical energy into the birth gas. Similar spatial distributions of the stars and gas after the shift of $(-10.15\degr, +1.44\degr)$ in $(\lmse,\bmse)$ indicate that \pwi birth cloud did not undergo significant gas removal and/or gas destruction period at all, or at least not at significant level. This might only be possible in the absence of stellar feedback in the \pwi birth cloud. 

One way to avoid the impact of stellar feedback on the gas cloud is not to have SN events. To test the possibility that no SNe occurred in \pwie, we compute the expected number of SN explosions in \pwie-like star clusters. Based on the present-day mass, age, and metallicity of \pwie, PARSEC stellar evolutionary models suggest that the initial mass of \pwi is $\sim$1800~\msune. We then simulate a 1800~\msun star cluster 20000 times assuming a Kroupa IMF and count the number of SN explosion events in each star cluster. If we assume that all stars more massive than 8~\msun explode as Type II SNe, then all of the simulated \pwie-like clusters produce at least 1 core-collapse SN $\sim$3 Myr (a typical lifetime of a 8~\msun star) after its birth. Thus, this scenario is unsuitable to explain the similar present-day spatial pattern between the \pwi and its birth cloud. 

Another way to avoid the gas disruption by stellar feedback is to spatially decouple the birth cloud and the newly formed stars before the first SN explosion. Over 3 Myr, the \pwi stars were able to travel $\sim$14.3 pc (corresponding to $\sim$50 lyr) away from the birth cloud based on the orbital calculation in \S \ref{subsec:orbit}.
This spatial decoupling due to ram pressure might prevent the \pwi birth cloud from being significantly disrupted by stellar feedback. If the \pwi stars and gas were indeed decoupled before the first SN explosion, our assumption about no effect of SN on the stellar motions (\S \ref{subsec:orbit}) can be naturally justified.  




\section{Summary} \label{sec:summary}

We have obtained high-resolution Magellan+MIKE spectra of 28 candidates of \pwie, a young stellar association in the region of the Leading Arm. Our sample allows us to draw some important conclusions about the properties and origin of both \pwi and the Leading Arm:

\begin{enumerate}
\item \pwi has a median metallicity of [Fe/H]=$-$1.23 with a small scatter of 0.06 dex and an inferred velocity of $\vlsre=\measvlsrk$~\kms with a dispersion of \measvdisp \kmse. The derived stellar parameters (\teffe, \logge, [Fe/H]) are consistent with the young, metal-poor isochrone (116 Myr and [Fe/H]=$-$1.1) that was determined in Paper I using photometry for proper-motion selected members.

\item There is a strong correlation between the spatial patterns of the \pwi stars and the high-density \hi clumps of \laii with an offset of ($-$10.15\degr,$+$1.55\degr) in (\lmse,\bmse) (\autoref{fig_shiftmap}).

\item Due to the similarity of metallicity, velocity, spatial patterns, and the distance of \pwie, we find that \pwi likely originated from the \laii complex of the Magellanic Stream.

\item The orbit and metallicity of \pwi and \laii associate them with the Magellanic Clouds and Magellanic Stream, in contrast to some recent claims to the contrary.

\item Using an orbital analysis of the \pwi stars and the \laii gas, taking into account ram pressure from a MW model, we constrain the halo gas density at the orbital pericenter of \pwi to be $n_{\rm halo}(17~\kpc) = 2.7_{-2.0}^{+3.4} \times 10^{-3}~{\rm atoms}~{\rm cm}^{-3}$ and the disk gas density at the midplane at $20~\kpc$ to be $n_{\rm disk}(20~\kpc, 0) = 6.0_{-2.0}^{+1.5} \times 10^{-2}~{\rm atoms}~{\rm cm}^{-3}$.  We also predict that the current distance of the \pwi birth cloud in \laii is 27 kpc.
\end{enumerate}

Future work will investigate the detailed chemical abundances of \pwi and how it compares to the Magellanic Clouds.

\acknowledgments
We thank Carlos Allende Prieto for sharing his Synspec IDL wrapper and utility software. We also thank Andy Casey for general help in running his Cannon software. We thank Michael Strauss for organizing and scheduling our Magellan time. APW thanks Stephanie Tonnesen and Elena D'Onghia for helpful conversations. DB and NWE thank Jane Luu for valuable advice and assistance in using the Goodman spectrograph at the SOAR telescope.  We thank Andrew Fox and Kat Barger for useful discussions on the origin of \pwi and the metallicity variations in the Leading Arm.
We thank the anonymous referee for useful comments that improved the manuscript.
Support for this work was provided by NASA through Hubble Fellowship grant \#51386.01 awarded to R.L.B. by the Space Telescope Science Institute, which is operated by the Association of  Universities for Research in Astronomy, Inc., for NASA, under contract NAS 5-26555.
This work presents results from the European Space Agency (ESA) space mission Gaia. Gaia data are being processed by the Gaia Data Processing and Analysis Consortium (DPAC). Funding for the DPAC is provided by national institutions, in particular the institutions participating in the Gaia MultiLateral Agreement (MLA). The Gaia mission website is https://www.cosmos.esa.int/gaia. The Gaia archive website is https://archives.esac.esa.int/gaia.
This paper includes data gathered with the 6.5 meter Magellan Telescopes located at Las Campanas Observatory, Chile. 

Based on observations obtained at the Southern Astrophysical Research (SOAR) telescope, which is a joint project of the Minist\'{e}rio da Ci\^{e}ncia, Tecnologia, Inova\c{c}\~{o}es e Comunica\c{c}\~{o}es (MCTIC) do Brasil, the U.S. National Optical Astronomy Observatory (NOAO), the University of North Carolina at Chapel Hill (UNC), and Michigan State University (MSU).

\facility{Magellan:Clay (MIKE)}
\facility{SOAR (Goodman)}
\software{
    CarPy: MIKE \citep{Kelson2000,Kelson2003},
    \package{Synspec} \citep{Hubeny2017},
    \package{IRAF} \citep{Tody1986, Tody1993},
    \package{The Cannon} \citep{Ness2017,Casey2012},
    \package{Astropy} \citep{astropy},
    \texttt{gala} \citep{PriceWhelan2017},
    \package{IPython} \citep{ipython},
    \package{matplotlib} \citep{mpl},
    \package{numpy} \citep{numpy},
    \package{scipy} \citep{scipy}
}

\bibliographystyle{aasjournals}
\bibliography{ref_og.bib}

\begin{thebibliography}{}
\expandafter\ifx\csname natexlab\endcsname\relax\def\natexlab#1{#1}\fi
\providecommand{\url}[1]{\href{#1}{#1}}

\bibitem[{{Bernstein} {et~al.}(2003){Bernstein}, {Shectman}, {Gunnels},
  {Mochnacki}, \& {Athey}}]{Bernstein2003}
{Bernstein}, R., {Shectman}, S.~A., {Gunnels}, S.~M., {Mochnacki}, S., \&
  {Athey}, A.~E. 2003, in \procspie, Vol. 4841, Instrument Design and
  Performance for Optical/Infrared Ground-based Telescopes, ed. M.~{Iye} \&
  A.~F.~M. {Moorwood}, 1694--1704

\bibitem[{{Besla} {et~al.}(2007){Besla}, {Kallivayalil}, {Hernquist},
  {Robertson}, {Cox}, {van der Marel}, \& {Alcock}}]{Besla2007}
{Besla}, G., {Kallivayalil}, N., {Hernquist}, L., {et~al.} 2007, \apj, 668, 949

\bibitem[{{Besla} {et~al.}(2012){Besla}, {Kallivayalil}, {Hernquist}, {van der
  Marel}, {Cox}, \& {Kere{\v s}}}]{Besla2012}
---. 2012, \mnras, 421, 2109

\bibitem[{Bovy(2015)}]{Bovy2015galpy}
Bovy, J. 2015, The Astrophysical Journal Supplement Series, 216, 29

\bibitem[{Brown {et~al.}(2018)Brown, Vallenari, Prusti, \&
  de~Bruijne}]{GaiaDR2}
Brown, A. G.~A., Vallenari, A., Prusti, T., \& de~Bruijne, J. H.~J. 2018,
  Astronomy {\&} Astrophysics, doi:10.1051/0004-6361/201833051

\bibitem[{{Brueck} \& {Hawkins}(1983)}]{Bruek1983}
{Brueck}, M.~T., \& {Hawkins}, M.~R.~S. 1983, \aap, 124, 216

\bibitem[{{Br{\"u}ns} {et~al.}(2005){Br{\"u}ns}, {Kerp}, {Staveley-Smith},
  {Mebold}, {Putman}, {Haynes}, {Kalberla}, {Muller}, \&
  {Filipovic}}]{Bruens2005}
{Br{\"u}ns}, C., {Kerp}, J., {Staveley-Smith}, L., {et~al.} 2005, \aap, 432, 45

\bibitem[{Byrd {et~al.}(1995)Byrd, Lu, Nocedal, \& Zhu}]{bfgs}
Byrd, R.~H., Lu, P., Nocedal, J., \& Zhu, C. 1995, SIAM J. Sci. Comput., 16,
  1190.
\newblock \url{http://dx.doi.org/10.1137/0916069}

\bibitem[{Carpenter {et~al.}(2017)Carpenter, Gelman, Hoffman, Lee, Goodrich,
  Betancourt, Brubaker, Guo, Li, \& Riddell}]{stan}
Carpenter, B., Gelman, A., Hoffman, M., {et~al.} 2017, Journal of Statistical
  Software, Articles, 76, 1.
\newblock \url{https://www.jstatsoft.org/v076/i01}

\bibitem[{{Casetti-Dinescu} {et~al.}(2014){Casetti-Dinescu}, {Moni Bidin},
  {Girard}, {M{\'e}ndez}, {Vieira}, {Korchagin}, \& {van
  Altena}}]{CasettiDinescu2014}
{Casetti-Dinescu}, D.~I., {Moni Bidin}, C., {Girard}, T.~M., {et~al.} 2014,
  \apjl, 784, L37

\bibitem[{{Casey} {et~al.}(2016){Casey}, {Hogg}, {Ness}, {Rix}, {Ho}, \&
  {Gilmore}}]{Casey2016}
{Casey}, A.~R., {Hogg}, D.~W., {Ness}, M., {et~al.} 2016, arXiv e-prints,
  arXiv:1603.03040

\bibitem[{{Casey} {et~al.}(2012){Casey}, {Keller}, \& {Da Costa}}]{Casey2012}
{Casey}, A.~R., {Keller}, S.~C., \& {Da Costa}, G. 2012, \aj, 143, 88

\bibitem[{{Choi} {et~al.}(2018{\natexlab{a}}){Choi}, {Nidever}, {Olsen},
  {Blum}, {Besla}, {Zaritsky}, {van der Marel}, {Bell}, {Gallart}, {Cioni},
  {Johnson}, {Vivas}, {Saha}, {de Boer}, {No{\"e}l}, {Monachesi}, {Massana},
  {Conn}, {Martinez-Delgado}, {Mu{\~n}oz}, \& {Stringfellow}}]{Choi2018a}
{Choi}, Y., {Nidever}, D.~L., {Olsen}, K., {et~al.} 2018{\natexlab{a}}, \apj,
  866, 90

\bibitem[{{Choi} {et~al.}(2018{\natexlab{b}}){Choi}, {Nidever}, {Olsen},
  {Besla}, {Blum}, {Zaritsky}, {Cioni}, {van der Marel}, {Bell}, {Johnson},
  {Vivas}, {Walker}, {de Boer}, {Noel}, {Monachesi}, {Gallart}, {Monelli},
  {Stringfellow}, {Massana}, \& {Martinez-Delgado}}]{Choi2018b}
---. 2018{\natexlab{b}}, ApJ, 869, 2

\bibitem[{{Clemens} {et~al.}(2004){Clemens}, {Crain}, \&
  {Anderson}}]{Clemens2004}
{Clemens}, J.~C., {Crain}, J.~A., \& {Anderson}, R. 2004, in Society of
  Photo-Optical Instrumentation Engineers (SPIE) Conference Series, Vol. 5492,
  \procspie, ed. A.~F.~M. {Moorwood} \& M.~{Iye}, 331--340

\bibitem[{{Connors} {et~al.}(2006){Connors}, {Kawata}, \&
  {Gibson}}]{Connors2006}
{Connors}, T.~W., {Kawata}, D., \& {Gibson}, B.~K. 2006, \mnras, 371, 108

\bibitem[{{Delhaye}(1965)}]{Delhaye1965}
{Delhaye}, J. 1965, {Solar Motion and Velocity Distribution of Common Stars},
  61

\bibitem[{{Diaz} \& {Bekki}(2012)}]{Diaz2012}
{Diaz}, J.~D., \& {Bekki}, K. 2012, \apj, 750, 36

\bibitem[{{D'Onghia} \& {Fox}(2016)}]{Donghia2016}
{D'Onghia}, E., \& {Fox}, A.~J. 2016, \araa, 54, 363

\bibitem[{{Dorman} {et~al.}(1993){Dorman}, {Rood}, \& {O'Connell}}]{Dorman1993}
{Dorman}, B., {Rood}, R.~T., \& {O'Connell}, R.~W. 1993, \apj, 419, 596

\bibitem[{Dormand \& Prince(1980)}]{Dormand1980}
Dormand, J., \& Prince, P. 1980, Journal of Computational and Applied
  Mathematics, 6, 19 .
\newblock
  \url{http://www.sciencedirect.com/science/article/pii/0771050X80900133}

\bibitem[{{Erkal} {et~al.}(2019){Erkal}, {Belokurov}, {Laporte}, {Koposov},
  {Li}, {Grillmair}, {Kallivayalil}, {Price-Whelan}, {Evans}, {Hawkins},
  {Hendel}, {Mateu}, {Navarro}, {del Pino}, {Slater}, {Sohn}, \& {Orphan Aspen
  Treasury Collaboration}}]{Erkal2019}
{Erkal}, D., {Belokurov}, V., {Laporte}, C.~F.~P., {et~al.} 2019, \mnras, 487,
  2685

\bibitem[{{For} {et~al.}(2013){For}, {Staveley-Smith}, \&
  {McClure-Griffiths}}]{For2013}
{For}, B.-Q., {Staveley-Smith}, L., \& {McClure-Griffiths}, N.~M. 2013, \apj,
  764, 74

\bibitem[{{Foreman-Mackey} {et~al.}(2013){Foreman-Mackey}, {Hogg}, {Lang}, \&
  {Goodman}}]{emcee}
{Foreman-Mackey}, D., {Hogg}, D.~W., {Lang}, D., \& {Goodman}, J. 2013,
  Publications of the Astronomical Society of the Pacific, 125, 306

\bibitem[{{Fox} {et~al.}(2013{\natexlab{a}}){Fox}, {Richter}, {Wakker},
  {Lehner}, {Howk}, {Ben Bekhti}, {Bland-Hawthorn}, \& {Lucas}}]{Fox2013a}
{Fox}, A.~J., {Richter}, P., {Wakker}, B.~P., {et~al.} 2013{\natexlab{a}},
  \apj, 772, 110

\bibitem[{{Fox} {et~al.}(2013{\natexlab{b}}){Fox}, {Richter}, {Wakker},
  {Lehner}, {Howk}, \& {Bland-Hawthorn}}]{Fox2013b}
---. 2013{\natexlab{b}}, The Messenger, 153, 28

\bibitem[{{Fox} {et~al.}(2010){Fox}, {Wakker}, {Smoker}, {Richter}, {Savage},
  \& {Sembach}}]{Fox2010}
{Fox}, A.~J., {Wakker}, B.~P., {Smoker}, J.~V., {et~al.} 2010, \apj, 718, 1046

\bibitem[{{Fox} {et~al.}(2018){Fox}, {Barger}, {Wakker}, {Richter},
  {Antwi-Danso}, {Casetti-Dinescu}, {Howk}, {Lehner}, {D'Onghia}, {Crowther},
  \& {Lockman}}]{Fox2018}
{Fox}, A.~J., {Barger}, K.~A., {Wakker}, B.~P., {et~al.} 2018, \apj, 854, 142

\bibitem[{{Gardiner} \& {Noguchi}(1996)}]{Gardiner1996}
{Gardiner}, L.~T., \& {Noguchi}, M. 1996, \mnras, 278, 191

\bibitem[{{Goodman} \& {Weare}(2010)}]{Goodman:2010}
{Goodman}, J., \& {Weare}, J. 2010, Communications in Applied Mathematics and
  Computational Science, 5, 65

\bibitem[{{Gordon}(1976)}]{Gordon1976}
{Gordon}, M.~A. 1976, Methods of Experimental Physics, 12, 277

\bibitem[{{Guhathakurta} \& {Reitzel}(1998)}]{Guhathakurta1998}
{Guhathakurta}, P., \& {Reitzel}, D.~B. 1998, in Astronomical Society of the
  Pacific Conference Series, Vol. 136, Galactic Halos, ed. D.~{Zaritsky}, 22

\bibitem[{{Harris} \& {Zaritsky}(2004)}]{Harris2004}
{Harris}, J., \& {Zaritsky}, D. 2004, \aj, 127, 1531

\bibitem[{{Harris} \& {Zaritsky}(2009)}]{Harris&Zaritsky2009}
---. 2009, \aj, 138, 1243

\bibitem[{Helmi {et~al.}(2018)Helmi, van Leeuwen, McMillan, \&
  DPAC}]{Helmi2018}
Helmi, A., van Leeuwen, F., McMillan, P., \& DPAC. 2018, Astronomy {\&}
  Astrophysics, doi:10.1051/0004-6361/201832698

\bibitem[{{Hernquist}(1990)}]{Hernquist1990}
{Hernquist}, L. 1990, \apj, 356, 359

\bibitem[{Homan \& Gelman(2014)}]{NUTS}
Homan, M.~D., \& Gelman, A. 2014, J. Mach. Learn. Res., 15, 1593.
\newblock \url{http://dl.acm.org/citation.cfm?id=2627435.2638586}

\bibitem[{{Hubeny} \& {Lanz}(2011)}]{Synspec}
{Hubeny}, I., \& {Lanz}, T. 2011, {Synspec: General Spectrum Synthesis
  Program}, , , ascl:1109.022

\bibitem[{{Hubeny} \& {Lanz}(2017)}]{Hubeny2017}
---. 2017, arXiv e-prints, arXiv:1706.01859

\bibitem[{{Hunter}(2007)}]{mpl}
{Hunter}, J.~D. 2007, Computing in Science and Engineering, 9, 90

\bibitem[{Jones {et~al.}(2001--)Jones, Oliphant, Peterson, {et~al.}}]{scipy}
Jones, E., Oliphant, T., Peterson, P., {et~al.} 2001--, {SciPy}: Open source
  scientific tools for {Python}, , .
\newblock \url{http://www.scipy.org/}

\bibitem[{{Kalberla} \& {Kerp}(2009)}]{Kalberla2009}
{Kalberla}, P. M.~W., \& {Kerp}, J. 2009, \araa, 47, 27

\bibitem[{{Kallivayalil} {et~al.}(2006){Kallivayalil}, {van der Marel},
  {Alcock}, {Axelrod}, {Cook}, {Drake}, \& {Geha}}]{Kallivayalil2006}
{Kallivayalil}, N., {van der Marel}, R.~P., {Alcock}, C., {et~al.} 2006, \apj,
  638, 772

\bibitem[{{Kallivayalil} {et~al.}(2013){Kallivayalil}, {van der Marel},
  {Besla}, {Anderson}, \& {Alcock}}]{Kallivayalil2013}
{Kallivayalil}, N., {van der Marel}, R.~P., {Besla}, G., {Anderson}, J., \&
  {Alcock}, C. 2013, \apj, 764, 161

\bibitem[{{Kelson}(2003)}]{Kelson2003}
{Kelson}, D.~D. 2003, \pasp, 115, 688

\bibitem[{{Kelson} {et~al.}(2000){Kelson}, {Illingworth}, {van Dokkum}, \&
  {Franx}}]{Kelson2000}
{Kelson}, D.~D., {Illingworth}, G.~D., {van Dokkum}, P.~G., \& {Franx}, M.
  2000, \apj, 531, 159

\bibitem[{{Kunkel} {et~al.}(1997){Kunkel}, {Irwin}, \& {Demers}}]{Kunkel1997}
{Kunkel}, W.~E., {Irwin}, M.~J., \& {Demers}, S. 1997, \aaps, 122, 463

\bibitem[{{Laporte} {et~al.}(2018){Laporte}, {G{\'o}mez}, {Besla}, {Johnston},
  \& {Garavito-Camargo}}]{Laporte2018b}
{Laporte}, C. F.~P., {G{\'o}mez}, F.~A., {Besla}, G., {Johnston}, K.~V., \&
  {Garavito-Camargo}, N. 2018, \mnras, 473, 1218

\bibitem[{{Lehner} {et~al.}(2008){Lehner}, {Howk}, {Keenan}, \&
  {Smoker}}]{Lehner2008}
{Lehner}, N., {Howk}, J.~C., {Keenan}, F.~P., \& {Smoker}, J.~V. 2008, \apj,
  678, 219

\bibitem[{{Lu} {et~al.}(1998){Lu}, {Savage}, {Sembach}, {Wakker}, {Sargent}, \&
  {Oosterloo}}]{Lu1998}
{Lu}, L., {Savage}, B.~D., {Sembach}, K.~R., {et~al.} 1998, \aj, 115, 162

\bibitem[{{Majewski} {et~al.}(2009){Majewski}, {Nidever}, {Mu{\~n}oz},
  {Patterson}, {Kunkel}, \& {Carlin}}]{Majewski2009}
{Majewski}, S.~R., {Nidever}, D.~L., {Mu{\~n}oz}, R.~R., {et~al.} 2009, in IAU
  Symposium, Vol. 256, The Magellanic System: Stars, Gas, and Galaxies, ed.
  J.~T. {Van Loon} \& J.~M. {Oliveira}, 51--56

\bibitem[{{Mastropietro} {et~al.}(2005){Mastropietro}, {Moore}, {Mayer},
  {Wadsley}, \& {Stadel}}]{Mastropietro2005}
{Mastropietro}, C., {Moore}, B., {Mayer}, L., {Wadsley}, J., \& {Stadel}, J.
  2005, \mnras, 363, 509

\bibitem[{{Mathewson} {et~al.}(1974){Mathewson}, {Cleary}, \&
  {Murray}}]{Mathewson1974}
{Mathewson}, D.~S., {Cleary}, M.~N., \& {Murray}, J.~D. 1974, \apj, 190, 291

\bibitem[{{McClure-Griffiths} {et~al.}(2008){McClure-Griffiths},
  {Staveley-Smith}, {Lockman}, {Calabretta}, {Ford}, {Kalberla}, {Murphy},
  {Nakanishi}, \& {Pisano}}]{McClure-Griffiths2008}
{McClure-Griffiths}, N.~M., {Staveley-Smith}, L., {Lockman}, F.~J., {et~al.}
  2008, \apjl, 673, L143

\bibitem[{{McClure-Griffiths} {et~al.}(2009){McClure-Griffiths}, {Pisano},
  {Calabretta}, {Ford}, {Lockman}, {Staveley-Smith}, {Kalberla}, {Bailin},
  {Dedes}, {Janowiecki}, {Gibson}, {Murphy}, {Nakanishi}, \&
  {Newton-McGee}}]{McClure-Griffiths2009}
{McClure-Griffiths}, N.~M., {Pisano}, D.~J., {Calabretta}, M.~R., {et~al.}
  2009, \apjs, 181, 398

\bibitem[{{Miller} \& {Bregman}(2013)}]{Miller2013}
{Miller}, M.~J., \& {Bregman}, J.~N. 2013, \apj, 770, 118

\bibitem[{{Mu{\~n}oz} {et~al.}(2006){Mu{\~n}oz}, {Majewski}, {Zaggia},
  {Kunkel}, {Frinchaboy}, {Nidever}, {Crnojevic}, {Patterson}, {Crane},
  {Johnston}, {Sohn}, {Bernstein}, \& {Shectman}}]{Munoz2006}
{Mu{\~n}oz}, R.~R., {Majewski}, S.~R., {Zaggia}, S., {et~al.} 2006, \apj, 649,
  201

\bibitem[{{Ness} {et~al.}(2015){Ness}, {Hogg}, {Rix}, {Ho}, \&
  {Zasowski}}]{Ness2015}
{Ness}, M., {Hogg}, D.~W., {Rix}, H.-W., {Ho}, A.~Y.~Q., \& {Zasowski}, G.
  2015, \apj, 808, 16

\bibitem[{{Ness} {et~al.}(2017){Ness}, {Rix}, {Hogg}, {Casey}, {Holtzman},
  {Fouesneau}, {Zasowski}, {Geisler}, {Shetrone}, {Minniti}, {Frinchaboy}, \&
  {Roman-Lopes}}]{Ness2017}
{Ness}, M., {Rix}, H., {Hogg}, D.~W., {et~al.} 2017, ArXiv e-prints,
  arXiv:1701.07829

\bibitem[{{Nidever} {et~al.}(2008){Nidever}, {Majewski}, \& {Butler
  Burton}}]{Nidever2008}
{Nidever}, D.~L., {Majewski}, S.~R., \& {Butler Burton}, W. 2008, \apj, 679,
  432

\bibitem[{{Nidever} {et~al.}(2010){Nidever}, {Majewski}, {Butler Burton}, \&
  {Nigra}}]{Nidever2010}
{Nidever}, D.~L., {Majewski}, S.~R., {Butler Burton}, W., \& {Nigra}, L. 2010,
  \apj, 723, 1618

\bibitem[{{Nidever} {et~al.}(2011){Nidever}, {Majewski}, {Mu{\~n}oz}, {Beaton},
  {Patterson}, \& {Kunkel}}]{Nidever2011}
{Nidever}, D.~L., {Majewski}, S.~R., {Mu{\~n}oz}, R.~R., {et~al.} 2011, \apjl,
  733, L10

\bibitem[{{Nidever} {et~al.}(2019{\natexlab{a}}){Nidever}, {Hasselquist},
  {Hayes}, {Hawkins}, {Majewski}, {Smith}, {Anguiano}, {Povick},
  {Stringfellow}, {Sobeck}, {Cunha}, {Nitschelm}, {Fernandez-Trincado},
  {Shetrone}, {Beers}, {Cohen}, {Allende Prieto}, {Gallart},
  {Garcia-Hernandez}, {Dell'Agli}, {Jonsson}, \& {Lacerna}}]{Nidever2019b}
{Nidever}, D.~L., {Hasselquist}, S., {Hayes}, C.~R., {et~al.}
  2019{\natexlab{a}}, arXiv e-prints, arXiv:1901.03448

\bibitem[{{Nidever} {et~al.}(2019{\natexlab{b}}){Nidever}, {Olsen}, {Choi}, {de
  Boer}, {Blum}, {Bell}, {Zaritsky}, {Martin}, {Saha}, {Conn}, {Besla}, {van
  der Marel}, {No{\"e}l}, {Monachesi}, {Stringfellow}, {Massana}, {Cioni},
  {Gallart}, {Monelli}, {Martinez-Delgado}, {Mu{\~n}oz}, {Majewski}, {Vivas},
  {Walker}, {Kaleida}, \& {Chu}}]{Nidever2019a}
{Nidever}, D.~L., {Olsen}, K., {Choi}, Y., {et~al.} 2019{\natexlab{b}}, \apj,
  874, 118

\bibitem[{{No{\"e}l} {et~al.}(2015){No{\"e}l}, {Conn}, {Read}, {Carrera},
  {Dolphin}, \& {Rix}}]{Noel2015}
{No{\"e}l}, N.~E.~D., {Conn}, B.~C., {Read}, J.~I., {et~al.} 2015, \mnras, 452,
  4222

\bibitem[{{Oh} {et~al.}(2017){Oh}, {Price-Whelan}, {Hogg}, {Morton}, \&
  {Spergel}}]{Oh2017}
{Oh}, S., {Price-Whelan}, A.~M., {Hogg}, D.~W., {Morton}, T.~D., \& {Spergel},
  D.~N. 2017, \aj, 153, 257

\bibitem[{{Olano}(2004)}]{Olano2004}
{Olano}, C.~A. 2004, \aap, 423, 895

\bibitem[{{Olsen} {et~al.}(2011){Olsen}, {Zaritsky}, {Blum}, {Boyer}, \&
  {Gordon}}]{Olsen2011}
{Olsen}, K.~A.~G., {Zaritsky}, D., {Blum}, R.~D., {Boyer}, M.~L., \& {Gordon},
  K.~D. 2011, \apj, 737, 29

\bibitem[{{Pagel} \& {Tautvaisiene}(1998)}]{Pagel1998}
{Pagel}, B.~E.~J., \& {Tautvaisiene}, G. 1998, \mnras, 299, 535

\bibitem[{{Pardy} {et~al.}(2018){Pardy}, {D'Onghia}, \& {Fox}}]{Pardy2018}
{Pardy}, S.~A., {D'Onghia}, E., \& {Fox}, A.~J. 2018, \apj, 857, 101

\bibitem[{{Patel} {et~al.}(2017){Patel}, {Besla}, \& {Sohn}}]{Patel2017}
{Patel}, E., {Besla}, G., \& {Sohn}, S.~T. 2017, \mnras, 464, 3825

\bibitem[{P\'erez \& Granger(2007)}]{ipython}
P\'erez, F., \& Granger, B.~E. 2007, Computing in Science and Engineering, 9,
  21.
\newblock \url{http://ipython.org}

\bibitem[{{Philip}(1976{\natexlab{a}})}]{Philip1976a}
{Philip}, A.~G.~D. 1976{\natexlab{a}}, in \baas, Vol.~8, 352

\bibitem[{{Philip}(1976{\natexlab{b}})}]{Philip1976b}
{Philip}, A.~G.~D. 1976{\natexlab{b}}, in \baas, Vol.~8, 532

\bibitem[{{Price-Whelan}(2017)}]{PriceWhelan2017}
{Price-Whelan}, A.~M. 2017, The Journal of Open Source Software, 2, 388

\bibitem[{{Price-Whelan} {et~al.}(2018){Price-Whelan}, {Nidever}, {Choi},
  {Schlafly}, {Morton}, {Koposov}, \& {Belokurov}}]{PriceWhelan2018}
{Price-Whelan}, A.~M., {Nidever}, D.~L., {Choi}, Y., {et~al.} 2018, arXiv
  e-prints, arXiv:1811.05991

\bibitem[{{Putman} {et~al.}(2012){Putman}, {Peek}, \& {Joung}}]{putman2012}
{Putman}, M.~E., {Peek}, J.~E.~G., \& {Joung}, M.~R. 2012, \araa, 50, 491

\bibitem[{{Putman} {et~al.}(2003){Putman}, {Staveley-Smith}, {Freeman},
  {Gibson}, \& {Barnes}}]{Putman2003}
{Putman}, M.~E., {Staveley-Smith}, L., {Freeman}, K.~C., {Gibson}, B.~K., \&
  {Barnes}, D.~G. 2003, \apj, 586, 170

\bibitem[{{Putman} {et~al.}(1998){Putman}, {Gibson}, {Staveley-Smith}, {Banks},
  {Barnes}, {Bhatal}, {Disney}, {Ekers}, {Freeman}, {Haynes}, {Henning},
  {Jerjen}, {Kilborn}, {Koribalski}, {Knezek}, {Malin}, {Mould}, {Oosterloo},
  {Price}, {Ryder}, {Sadler}, {Stewart}, {Stootman}, {Vaile}, {Webster}, \&
  {Wright}}]{Putman1998}
{Putman}, M.~E., {Gibson}, B.~K., {Staveley-Smith}, L., {et~al.} 1998, \nat,
  394, 752

\bibitem[{{Recillas-Cruz}(1982)}]{Recillas-Cruz1982}
{Recillas-Cruz}, E. 1982, \mnras, 201, 473

\bibitem[{{Richter} {et~al.}(2018){Richter}, {Fox}, {Wakker}, {Howk}, {Lehner},
  {Barger}, {D'Onghia}, \& {Lockman}}]{Richter2018}
{Richter}, P., {Fox}, A.~J., {Wakker}, B.~P., {et~al.} 2018, \apj, 865, 145

\bibitem[{{Richter} {et~al.}(2013){Richter}, {Fox}, {Wakker}, {Lehner}, {Howk},
  {Bland -Hawthorn}, {Ben Bekhti}, \& {Fechner}}]{Richter2013}
---. 2013, \apj, 772, 111

\bibitem[{{Russell} \& {Dopita}(1992)}]{Russell1992}
{Russell}, S.~C., \& {Dopita}, M.~A. 1992, \apj, 384, 508

\bibitem[{{Sch{\"o}nrich} {et~al.}(2010){Sch{\"o}nrich}, {Binney}, \&
  {Dehnen}}]{Schoenrich2010}
{Sch{\"o}nrich}, R., {Binney}, J., \& {Dehnen}, W. 2010, \mnras, 403, 1829

\bibitem[{{Stanimirovi{\'c}} {et~al.}(2008){Stanimirovi{\'c}}, {Hoffman},
  {Heiles}, {Douglas}, {Putman}, \& {Peek}}]{Stanimirovic2008}
{Stanimirovi{\'c}}, S., {Hoffman}, S., {Heiles}, C., {et~al.} 2008, \apj, 680,
  276

\bibitem[{{Tepper-Garc{\'{\i}}a} \& {Bland-Hawthorn}(2018)}]{Tepper-Garcia2018}
{Tepper-Garc{\'{\i}}a}, T., \& {Bland-Hawthorn}, J. 2018, \mnras, 478, 5263

\bibitem[{{The Astropy Collaboration} {et~al.}(2018){The Astropy
  Collaboration}, {Price-Whelan}, {Sip{\H o}cz}, {G{\"u}nther}, {Lim},
  {Crawford}, {Conseil}, {Shupe}, {Craig}, {Dencheva}, {Ginsburg},
  {VanderPlas}, {Bradley}, {P{\'e}rez-Su{\'a}rez}, {de Val-Borro}, {Aldcroft},
  {Cruz}, {Robitaille}, {Tollerud}, {Ardelean}, {Babej}, {Bachetti}, {Bakanov},
  {Bamford}, {Barentsen}, {Barmby}, {Baumbach}, {Berry}, {Biscani}, {Boquien},
  {Bostroem}, {Bouma}, {Brammer}, {Bray}, {Breytenbach}, {Buddelmeijer},
  {Burke}, {Calderone}, {Cano Rodr{\'{\i}}guez}, {Cara}, {Cardoso},
  {Cheedella}, {Copin}, {Crichton}, {D{\'A}vella}, {Deil}, {Depagne},
  {Dietrich}, {Donath}, {Droettboom}, {Earl}, {Erben}, {Fabbro}, {Ferreira},
  {Finethy}, {Fox}, {Garrison}, {Gibbons}, {Goldstein}, {Gommers}, {Greco},
  {Greenfield}, {Groener}, {Grollier}, {Hagen}, {Hirst}, {Homeier}, {Horton},
  {Hosseinzadeh}, {Hu}, {Hunkeler}, {Ivezi{\'c}}, {Jain}, {Jenness}, {Kanarek},
  {Kendrew}, {Kern}, {Kerzendorf}, {Khvalko}, {King}, {Kirkby}, {Kulkarni},
  {Kumar}, {Lee}, {Lenz}, {Littlefair}, {Ma}, {Macleod}, {Mastropietro},
  {McCully}, {Montagnac}, {Morris}, {Mueller}, {Mumford}, {Muna}, {Murphy},
  {Nelson}, {Nguyen}, {Ninan}, {N{\"o}the}, {Ogaz}, {Oh}, {Parejko}, {Parley},
  {Pascual}, {Patil}, {Patil}, {Plunkett}, {Prochaska}, {Rastogi}, {Reddy
  Janga}, {Sabater}, {Sakurikar}, {Seifert}, {Sherbert}, {Sherwood-Taylor},
  {Shih}, {Sick}, {Silbiger}, {Singanamalla}, {Singer}, {Sladen}, {Sooley},
  {Sornarajah}, {Streicher}, {Teuben}, {Thomas}, {Tremblay}, {Turner},
  {Terr{\'o}n}, {van Kerkwijk}, {de la Vega}, {Watkins}, {Weaver}, {Whitmore},
  {Woillez}, \& {Zabalza}}]{astropy}
{The Astropy Collaboration}, {Price-Whelan}, A.~M., {Sip{\H o}cz}, B.~M.,
  {et~al.} 2018, ArXiv e-prints, arXiv:1801.02634

\bibitem[{{Tody}(1986)}]{Tody1986}
{Tody}, D. 1986, in Society of Photo-Optical Instrumentation Engineers (SPIE)
  Conference Series, Vol. 627, \procspie, ed. D.~L. {Crawford}, 733

\bibitem[{{Tody}(1993)}]{Tody1993}
{Tody}, D. 1993, in Astronomical Society of the Pacific Conference Series,
  Vol.~52, Astronomical Data Analysis Software and Systems II, ed. R.~J.
  {Hanisch}, R.~J.~V. {Brissenden}, \& J.~{Barnes}, 173

\bibitem[{{Vollmer} {et~al.}(2001){Vollmer}, {Cayatte}, {Balkowski}, \&
  {Duschl}}]{Vollmer2001}
{Vollmer}, B., {Cayatte}, V., {Balkowski}, C., \& {Duschl}, W.~J. 2001, \apj,
  561, 708

\bibitem[{{Wakker}(2001)}]{Wakker2001}
{Wakker}, B.~P. 2001, \apjs, 136, 463

\bibitem[{Walt {et~al.}(2011)Walt, Colbert, \& Varoquaux}]{numpy}
Walt, S. v.~d., Colbert, S.~C., \& Varoquaux, G. 2011, Computing in Science and
  Engg., 13, 22.
\newblock \url{http://dx.doi.org/10.1109/MCSE.2011.37}

\bibitem[{{Wannier} {et~al.}(1972){Wannier}, {Wrixon}, \& {Wilson}}]{WW1972}
{Wannier}, P., {Wrixon}, G.~T., \& {Wilson}, R.~W. 1972, \aap, 18, 224

\bibitem[{{Weisz} {et~al.}(2013){Weisz}, {Dolphin}, {Skillman}, {Holtzman},
  {Dalcanton}, {Cole}, \& {Neary}}]{Weisz2013}
{Weisz}, D.~R., {Dolphin}, A.~E., {Skillman}, E.~D., {et~al.} 2013, \mnras,
  431, 364

\bibitem[{{Wong} {et~al.}(2011){Wong}, {Hughes}, {Ott}, {Muller}, {Pineda},
  {Bernard}, {Chu}, {Fukui}, {Gruendl}, {Henkel}, {Kawamura}, {Klein},
  {Looney}, {Maddison}, {Mizuno}, {Paradis}, {Seale}, \& {Welty}}]{Wong2011}
{Wong}, T., {Hughes}, A., {Ott}, J., {et~al.} 2011, \apjs, 197, 16

\bibitem[{{Zhang} {et~al.}(2019){Zhang}, {Casetti-Dinescu}, {Moni Bidin},
  {M{\'e}ndez}, {Girard}, {Vieira}, {Korchagin}, {van Altena}, \&
  {Zhao}}]{Zhang2019}
{Zhang}, L., {Casetti-Dinescu}, D.~I., {Moni Bidin}, C., {et~al.} 2019, \apj,
  871, 99

\bibitem[{{Zivick} {et~al.}(2018){Zivick}, {Kallivayalil}, {van der Marel},
  {Besla}, {Linden}, {Koz{\l}owski}, {Fritz}, {Kochanek}, {Anderson}, {Sohn},
  {Geha}, \& {Alcock}}]{Zivick2018}
{Zivick}, P., {Kallivayalil}, N., {van der Marel}, R.~P., {et~al.} 2018, \apj,
  864, 55

\end{thebibliography}


\clearpage
\movetabledown=2.0in
\begin{rotatetable}
\begin{deluxetable*}{lcccccccccccccccccc}
\tablecaption{\pwi Spectroscopic Results}\label{table1}
\tablecolumns{18}
\tablewidth{0pt}
\tablehead{
\colhead{Name} & \colhead{Gaia ID} & \colhead{RA} & \colhead{DEC} &   
\colhead{G} & \colhead{G$_{\rm BP}$-G$_{\rm RP}$} &
\colhead{$\mu_{\rm RA}$} & \colhead{$\mu_{\rm DEC}$} & 
\colhead{S/N} & \colhead{$V_{\rm LSR}$} & \colhead{$\sigma_V$} &
\colhead{\teffe} & \colhead{$\sigma_{\rm T_{eff}}$} &
\colhead{\logge} & \colhead{$\sigma_{\log{g}}$} &
\colhead{[Fe/H]} & \colhead{$\sigma_{\rm [Fe/H]}$} \\
\colhead{ } & \multicolumn2c{(J2000)} & \colhead{} & 
\multicolumn2c{(mag)} & \multicolumn2c{(mas yr$^{-1}$)} & 
\colhead{ } & \multicolumn2c{(km s$^{-1}$)} & \multicolumn2c{(K)} &
\multicolumn2c{(dex)} & \multicolumn2c{(dex)}
}
\startdata
PW1-00 & 3466763224691512064 & 11:55:31.8 & $-$33:09:11.6 & 15.10 & $-$0.11 & $-$0.58 & 0.46 & 11.8 & 239.7 & 7.3 & 14465 & 766 & 2.25 & 0.12 & $-$0.81 & 0.32 \\ 
PW1-01 & 3480054567924428032 & 11:56:00.4 & $-$29:28:48.9 & 16.30 & $-$0.21 & $-$0.45 & 0.42 & 14.2 & 257.1 & 6.7 & 17326 & 873 & 3.38 & 0.13 & $-$0.82 & 0.32 \\ 
PW1-02 & 3480064910205689088 & 11:56:18.3 & $-$29:19:20.8 & 16.39 & $-$0.24 & $-$0.85 & 0.55 & 12.4 & 255.3 & 7.3 & 16340 & 1000 & 3.20 & 0.16 & $-$1.06 & 0.35 \\ 
PW1-03 & 3480053399693321984 & 11:56:12.0 & $-$29:31:44.4 & 16.71 & $-$0.23 & $-$0.37 & 0.50 & 4.3 & 298.5 & 20.0 & 17436 & 2041 & 3.22 & 0.36 & $-$0.33 & 0.57 \\ 
PW1-04 & 3486242378847130624 & 12:00:11.5 & $-$27:59:50.4 & 16.78 & $-$0.17 & $-$0.76 & 0.45 & 8.8 & 235.1 & 11.3 & 14654 & 953 & 3.38 & 0.19 & $-$1.11 & 0.30 \\ 
PW1-05 & 3479870124848931200 & 11:57:20.8 & $-$29:27:45.2 & 16.84 & $-$0.20 & $-$0.58 & 0.55 & 16.1 & 262.7 & 5.3 & 17534 & 548 & 3.76 & 0.09 & $-$1.04 & 0.18 \\ 
PW1-06 & 3479874106283037568 & 11:56:56.3 & $-$29:25:11.6 & 16.84 & $-$0.20 & $-$0.71 & 0.57 & 29.1 & 259.0 & 3.3 & 15567 & 263 & 3.77 & 0.05 & $-$1.25 & 0.12 \\ 
PW1-07 & 3480052712498583424 & 11:55:46.3 & $-$29:33:56.9 & 16.87 & $-$0.24 & $-$0.43 & 0.50 & 30.6 & 267.9 & 2.7 & 17321 & 315 & 3.51 & 0.05 & $-$1.26 & 0.12 \\ 
PW1-08 & 3486167027940903296 & 11:56:04.8 & $-$28:28:38.8 & 16.95 & $-$0.21 & $-$0.47 & 0.59 & 32.0 & 274.9 & 2.7 & 17930 & 321 & 3.64 & 0.05 & $-$1.32 & 0.10 \\ 
PW1-09 & 3480071644714348032 & 11:55:41.7 & $-$29:17:32.5 & 17.01 & $-$0.23 & $-$0.53 & 0.46 & 29.3 & 266.9 & 3.3 & 17878 & 350 & 3.62 & 0.05 & $-$1.24 & 0.12 \\ 
PW1-10 & 3479874690398598784 & 11:57:09.5 & $-$29:21:50.6 & 17.09 & $-$0.18 & $-$0.54 & 0.40 & 27.2 & 288.7 & 4.0 & 15805 & 296 & 3.64 & 0.06 & $-$1.16 & 0.14 \\ 
PW1-11 & 3485879024613971328 & 11:57:42.5 & $-$29:20:32.8 & 17.14 & $-$0.12 & 0.22 & 0.42 & 22.7 & 278.2 & 4.7 & 14561 & 329 & 3.60 & 0.07 & $-$1.17 & 0.15 \\ 
PW1-12 & 3480036975738300800 & 11:53:53.6 & $-$29:38:55.3 & 17.17 & $-$0.17 & $-$0.49 & 0.47 & 23.4 & 262.4 & 4.0 & 16105 & 372 & 3.66 & 0.07 & $-$1.25 & 0.14 \\ 
PW1-13 & 3467765189021964544 & 12:00:18.7 & $-$30:14:30.8 & 17.19 & $-$0.16 & $-$0.23 & 0.30 & 19.9 & 341.6 & 5.3 & 17216 & 451 & 4.19 & 0.07 & $-$1.15 & 0.17 \\ 
PW1-14 & 3480121844292017920 & 11:54:23.4 & $-$29:15:42.0 & 17.42 & $-$0.22 & $-$0.40 & 0.44 & 24.2 & 280.7 & 4.0 & 17254 & 401 & 3.78 & 0.06 & $-$1.24 & 0.14 \\ 
PW1-15 & 3480049757560914944 & 11:54:25.0 & $-$29:23:44.4 & 17.54 & $-$0.20 & $-$0.50 & 0.21 & 32.9 & 267.2 & 2.7 & 17472 & 283 & 3.67 & 0.05 & $-$1.18 & 0.09 \\ 
PW1-16 & 3480046557809199616 & 11:54:46.1 & $-$29:25:33.5 & 17.57 & $-$0.22 & $-$0.54 & 0.45 & 37.7 & 266.4 & 2.7 & 15726 & 225 & 3.31 & 0.04 & $-$1.16 & 0.11 \\ 
PW1-17 & 3480070957519575552 & 11:55:27.6 & $-$29:21:43.8 & 17.59 & $-$0.17 & $-$0.48 & 0.27 & 37.8 & 268.3 & 2.7 & 16894 & 235 & 3.66 & 0.04 & $-$1.27 & 0.08 \\ 
PW1-18 & 3479762991184812544 & 11:57:32.4 & $-$30:14:56.6 & 17.67 & $-$0.19 & $-$1.00 & 0.36 & 38.1 & 275.9 & 2.7 & 15871 & 218 & 3.62 & 0.04 & $-$1.25 & 0.08 \\ 
PW1-19 & 3486175647939287680 & 11:57:29.7 & $-$28:29:54.7 & 17.70 & $-$0.12 & $-$0.22 & 0.29 & 31.5 & 257.7 & 2.7 & 13979 & 224 & 3.49 & 0.05 & $-$1.20 & 0.10 \\ 
PW1-20 & 3479873354664338688 & 11:57:09.4 & $-$29:25:47.3 & 17.69 & $-$0.07 & $-$0.61 & 0.37 & 29.8 & 244.3 & 4.0 & 13609 & 210 & 3.82 & 0.05 & $-$1.22 & 0.10 \\ 
PW1-21 & 3486267147922619136 & 12:00:22.2 & $-$27:56:54.3 & 17.74 & $-$0.18 & $-$0.43 & 0.18 & 31.6 & 289.6 & 2.7 & 16060 & 303 & 3.38 & 0.05 & $-$1.28 & 0.11 \\ 
PW1-22 & 3480047141924659456 & 11:53:57.4 & $-$29:30:37.4 & 17.75 & $-$0.18 & $-$0.40 & 0.40 & 28.5 & 277.2 & 3.3 & 15402 & 280 & 3.73 & 0.05 & $-$1.28 & 0.11 \\ 
PW1-23 & 3480049856344140288 & 11:54:37.3 & $-$29:22:36.4 & 17.80 & $-$0.06 & $-$0.53 & 0.61 & 31.9 & 284.5 & 2.7 & 15845 & 285 & 3.35 & 0.05 & $-$1.23 & 0.12 \\ 
PW1-24 & 3485874897149344256 & 11:57:58.3 & $-$29:24:47.2 & 17.79 & $-$0.12 & $-$0.51 & 0.44 & 29.9 & 257.2 & 3.3 & 15246 & 291 & 3.47 & 0.06 & $-$1.19 & 0.12 \\ 
PW1-25 & 3480070510842975104 & 11:55:07.5 & $-$29:21:14.9 & 17.82 & $-$0.08 & $-$0.71 & 0.37 & 29.3 & 271.2 & 4.0 & 13570 & 201 & 4.00 & 0.05 & $-$1.15 & 0.10 \\ 
PW1-26 & 3480122325328358144 & 11:54:25.7 & $-$29:13:24.8 & 18.30 & $-$0.23 & $-$0.42 & 0.45 & 21.3 & 292.6 & 4.7 & 14914 & 189 & 3.01 & 0.05 & $-$1.21 & 0.02 \\ 
PW1-27 & 3486267251001844736 & 12:00:23.1 & $-$27:55:21.8 & 17.94 & $-$0.12 & $-$0.89 & 0.37 & 29.3 & 270.0 & 3.3 & 15050 & 264 & 3.30 & 0.05 & $-$1.18 & 0.05
\enddata
\end{deluxetable*}
\end{rotatetable}

\end{document}